# The Theory of Uncertainty for Derived Results: Properties of Equations Representing Physicochemical Evaluation Systems

*B. P. Datta* (EMAIL: bibek@vecc.gov.in)

*Radiochemistry Laboratory, Variable Energy Cyclotron Centre, 1/AF Bidhan Nagar, Kolkata-700064, India*

## ABSTRACT

Any physiochemical variable ($Y_m$) is always determined from certain measured variables $\{X_i\}$. The uncertainties $\{u_i\}$ of measuring $\{X_i\}$ are generally a priori ensured as acceptable. However, there is no general method for assessing uncertainty ($\varepsilon_m$) in the desired $Y_m$, i.e. irrespective of whatever might be its system-specific-relationship (SSR) with $\{X_i\}$, and/ or be the causes of $\{u_i\}$. We here therefore study the behaviors of different typical SSRs. The study shows that any SSR is characterized by a set of parameters, which govern $\varepsilon_m$. That is, $\varepsilon_m$ is shown to represent a net SSR-driven (purely systematic) change in $u_i(s)$; and it cannot vary for whether $u_i(s)$ be caused by either or both statistical and systematic reasons. We thus present the general relationship of $\varepsilon_m$ with $u_i(s)$, and discuss how it can be used to predict a priori the requirements for an evaluated $Y_m$ to be representative, and hence to set the guidelines for designing experiments and also really appropriate evaluation models. Say: $Y_m = f_m(\{X_i\}_{i=1}^{N})$, then, although: $\varepsilon_m = g_m(\{u_i\}_{i=1}^{N})$, "$N$" is not a key factor in governing $\varepsilon_m$. However, simply for varying "$f_m$", the $\varepsilon_m$ is demonstrated to be either equaling a $u_i$, or $>u_i$, or even $<u_i$. Further, the limiting error ($\delta_m^{\text{Lim.}}$) in determining an $Y_m$ is also shown to be decided by "$f_m$" (SSR). Thus, all SSRs are classified into two groups: (I) the SSRs that can never lead "$\delta_m^{\text{Lim.}}$" to be zero; and (II) the SSRs that enable "$\delta_m^{\text{Lim.}}$" to be zero. In fact, the theoretical-tool (SSR) is by pros and cons no different from any discrete experimental-means of a study, and has resemblance with chemical reactions as well.

## 1. INTRODUCTION

Generally, no real world variable (viz. a parameter, or simply concentration, of a chemical species), $Y_m$, can be measured directly. That is the value of $Y_m$ is derived [1] from certain relevant measured variable(s), $X_i(s)$, by using their given system-specific-relationship (SSR):

$$Y_m = f_m(X_i), \qquad (i = m = N = 1) \tag{1a}$$

Or,

$$Y_m = f_m(\{X_i\}), \quad i = 1, 2, \ldots, N \text{ (for a given } m) \tag{1b}$$

That is, in reality, the desired result ($y_m$) is obtained as:

$$y_m = f_m(x_i) = f_m(X_i + \Delta_i), \qquad (i = m = N = 1) \tag{1a$^/$}$$

Or,

$$y_m = f_m(\{x_i\}) = f_m(\{X_i + \Delta_i\}), \qquad i = 1, 2, \ldots, N \text{ (for a given } m) \tag{1b$^/$}$$

where $\Delta_i$ stands for the deviation in the measured estimate $x_i$ from its unknown true value ($X_i$). Further, by Eq. 1a and Eq. 1b, we refer to here all conceivable ***individual*** evaluations as $Y_m$ ($m$ = 1, 2 …), for involving one ($i = N = 1$) [2,3] and more [4-7] than one ($i$ = 1, 2, … $N$) measured variable ($X_i$), respectively. However, in many a case [8-10], different $\{Y_m\}$ are determined ***simultaneously***, i.e. the evaluation is represented by a set of SSRs (equations):

$$Y_m = f_m(\{X_i\}), \qquad i, m = 1, 2, \ldots, N \tag{1c}$$

Or, in terms of the desired estimates $\{y_m\}$:

$$y_m = f_m(\{x_i\}) = f_m(\{X_i + \Delta_i\}), \qquad i, m = 1, 2, \ldots, N \tag{1c$^/$}$$

Anyway, the purpose of any evaluation [2-10] is to ascertain the corresponding $Y_m$-value(s). However, as indicated by Eq. $1^/$**:** $y_m = (Y_m + \delta_m)$, where $\delta_m$ stands for the error in $y_m$. Further, the experimental error(s) $\Delta_i(s)$, and thus $\delta_m$, cannot be known. Therefore, the question is raised here how the result $y_m$ is in any given case ensured to be representative of the desired $Y_m$.

It may however be pointed out that, unless at least the *highest possible value* (HPV) of the error $\Delta_i$ could be known, the data $x_i$ itself cannot be used. Again, the result-shaping (Eq. $1^{/}$) is a theoretical task. Thus, by requirements for an $y_m$ to be accurate, it may be meant the selection and development of simply $X_i$-measurement technique(s) such that the HPV(s) of experimental error(s) $\Delta_i$*(s)* are at least acceptably small. In support, it may be added that the result $y_m$ is usually considered valid if and when the variations in the corresponding measured estimates $\{x_i\}$ are acceptable [5,6]. We here denote the "*HPV of error in* $x_i$" by $u_i$ (i.e.: $u_i = |\Delta_i^{\mathrm{Max}}|$), and "*that in* $y_m$" by $\varepsilon_m$ (i.e.: $\varepsilon_m = |\delta_m^{\mathrm{Max}}|$), irrespective of whatever might be the relationship of $\varepsilon_m$ with $u_i$*(s)*. In any case, the values of (**method- and/ or**) $\boldsymbol{X_i}$**-specific** $u_i$ and $\boldsymbol{Y_m}$**-specific** $\varepsilon_m$ should signify how worst "$x_i$" and "$y_m$" might be deviating from "$X_i$" and "$Y_m$", respectively. Therefore, we refer the *HPV of error* as either *uncertainty* or *inaccuracy* (*accuracy*). However, the true measure of any error is always its relative value [11]. So, we define: $\Delta_i = \frac{\Delta X_i}{X_i} = \frac{x_i - X_i}{X_i}$ and: $\delta_m = \frac{\delta Y_m}{Y_m} = \frac{y_m - Y_m}{Y_m}$, and hence "$u_i$" and "$\varepsilon_m$" as the relative uncertainties (see also APPENDIX 1).

Further, by method-development, it should usually mean that the $X_i$-measurement is ensured to be bias-free. Thus, the *standard deviation* ($\sigma_i$) of repetitive measurements [1] should be the best estimate for the method cum $X_i$ -specific *HPV of error* ($u_i$). That is, it is generally expected that: $u_i = \sigma_i$. Again, unless $x_i$ is at error, $y_m$ can never be at error (cf. Eq. 1). These might explain why the result $y_m$ was sometimes reported without clarifying how well it represents the desired $Y_m$, and/ or why any corresponding measured data $x_i$ was usually presented [5,6] along with its scatter $\sigma_i$. Yet, it is here enquired whether (even for the simple type of systems [2,3] as Eq. 1a) "$u_i = \sigma_i$" can cause the resultant-uncertainty $\varepsilon_m$ to equal the measurement-uncertainty $u_i$. That is, a purpose here is to evaluate whether the result $y_m$ can ever turn out more uncertain (less accurate**:** $\varepsilon_m > u_i$), or even **less** uncertain (**more** accurate**:** $\varepsilon_m < u_i$), than the measured data $x_i$, and hence to show how the experimental goal as "$u_i$" to be achieved can a priori be preset.

Nevertheless, the desired result ($y_m$) is also used to be validated by the observed[4] or predicted scatter ($\rho_m$) of its own. The predicted value is called as the combined standard [1], or the probable cum propagated [11], uncertainty. The same is, for any given case of Eq. 1b and for all $\{X_i\}$ to be independent, computed here as:

$$\rho_m = \frac{1}{y_m}\left[\sum_{i=1}^{N}\left(\frac{\partial Y_m}{\partial X_i}\right)^2 \sigma_i^2\right]^{1/2} \qquad (2)$$

It may now be reminded that the *possible,* and hence the *highest* and *unaccountable*, error ($u_i$) in a measurement was in fact generally referred to as the uncertainty (cf. Section 0.2 in [1]). However, the uncertainty was recommended [1] to be measured in terms of standard deviation ($\sigma_i$) only, and implied to be **different** from inaccuracy. Of course, only a true value (*truth*) could be meant as 100% accurate (the *certainty*). Yet, it may be mentioned that *any* real world fact is truly subjective. We thus stick only to the basic concept that the ***uncertainty*** in a given measured estimate $\boldsymbol{x_i}$ should indicate whether or not "$\boldsymbol{x_i}$" is a ***good*** (*better*) representative of its true value $\boldsymbol{X_i}$ (*than* the estimate obtained by some other technique). That is there should be no alternative to considering experiment cum $X_i$-specific HPV of error ($u_i$) as its ***measure***, irrespective of whether "$u_i = \sigma_i$" or "$u_i$" is represented even by case-specific (*unidentified*/ *uncorrectable*) biases.

It may also be taken to note that, as "$\sigma_i$" for the uncertainty "$u_i$", Eq. 2 (or an appropriate form of it, representing a specific system of non-linear SSR(s) and/ or interdependency in $\{X_i\}$) was considered [1] to offer the best measure for the uncertainty ($\varepsilon_m$) in a corresponding derived result $y_m$, and was used [1,12-17] as the key to related developments. However we enquire whether really even "$\boldsymbol{u_i} = \boldsymbol{\sigma_i}$ ($\boldsymbol{i}$ = 1, 2 …, $\boldsymbol{N}$)" can cause "$\boldsymbol{\varepsilon_m} = \boldsymbol{\rho_m}$". Further, is "$\boldsymbol{N}$" a critical factor?

Over and above, the data $x_i$ was assumed [1] to be corrected for all possible systematic effects. However, the systematic effects may not get distinguished from the random ones, specifically in case of an intricate measurement. Or, it may in a given measurement happen that the systematic and the random effects are equally insignificant. However, even in that case**:** $u_i \neq \sigma_i$, and hence**:** $\varepsilon_m \neq \rho_m$. It was of course also suggested [1] how, even for such a case, the output uncertainty could be evaluated as $\rho_m$. Yet, we ask (cf. Eq. 1)**:** can, depending on whether the **causes** for the error in a given $x_i$ are purely statistical or systematic or both, $y_m$ vary? —— Thus, we point out that the result-shaping (cf. Eq. 1)**:** $x_i(s) \rightarrow y_m(s)$ itself stands for the SSR dictated (i.e. *desired*) **biasing** of the *given* data $x_i(s)$. That is, even when $x_i(s)$ should be at purely ***random*** error(s), the error in $y_m$ will by origin be ***systematic*** only. In fact, for any given equal but opposite errors in $x_i$, the errors in $y_m$ should (but depending upon the SSR, Eq. 1) be taking asymmetrical values. That is, in principle, the uncertainty in $y_m$ cannot be ascertained by any statistical-cum-distribution means [1,12-17], and hence the idea here is to look for the right method.

However, it could be best to approach the problem by elaborating our considerations on a real world case, viz. the evaluation [2] of a constituent elemental isotopic abundance ratio ($Y_m$) from the measured abundance ratio ($X_i$) of an isotopic $Li_2BO_2^+$ ion-pair (***i***)**:** $\boldsymbol{Y_m = f_m(X_i)}$. The measurement is carried out on certain isotopic *molecular*-ions, whereas the result is required for their constituent *elemental* isotopes. This explains why at all a theoretical task (Eq. 1a: $\boldsymbol{(y_m \pm \varepsilon_m) = f_m(x_i \pm u_i)}$) is involved in the study [2], which is basically an experimental one. However, as only the measured estimate $x_i$ should be subject to certain uncertainty ($u_i$), it is the idea here to evaluate why an SSR as even Eq. **1a** may cause the uncertainty $\varepsilon_m$ to vary from $u_i$. Thus, suppose, the use of relevant *standards* had clarified that**: (i)** at worst**:** $x_i = (X_i \pm 0.001X_i)$, i.e.**:** $u_i = 0.1\%$; **(ii)** $u_i$ is insensitive towards '$|\boldsymbol{X_i}|$'; and**: (iii)** $u_i$ is independent of '***i***' (i.e. say**:** $u_J = u_K$). Then, **(1)** should we believe, but why or why not, that**:** $\varepsilon_m = 0.01\%$? **(2)** Should $\varepsilon_m$ be invariant, like $u_i$, of '$|\boldsymbol{X_i}|$'? **(3)** Should, e.g. "$u_{55/57} = u_{56/57}$" ensure the result $y_m$ to be equally reliable for *selecting* the "$Li_2BO_2^+$" ion-pair of either m/z (55, 57) or m/z (56, 57) as the *monitors*? That is, can the function "$\boldsymbol{f_m(X_i)}$" have a say in **even** experimental planning? —— Further, should the results of considerations**: (1)-(3)** vary for whether $\boldsymbol{Y_m}$ stands for, e.g. $^6Li/^7Li$ abundance ratio ($\boldsymbol{Y_{Li}}$), or $^{10}B/^{11}B$ ratio ($\boldsymbol{Y_B}$), or so? Should $\varepsilon_B$ equal to $\varepsilon_{Li}$ and/ or $u_i$? —— Suppose further that**:** (i) the error ($\Delta_i$) and the standard deviation ($\sigma_i$) of measurements (on $X_i$-standards) did **never** exceed the acceptable limit (**0.1%**), but (ii) the method was so intricate that the causes of errors could **not** be ascertained. However, should $\varepsilon_m$ vary for **why** "$u_i$ = **0.1%**"?

Similarly a typical **two** variable (Eq. **1b**) system is the determination of OH-acetone reaction rate constant ($Y_R$) by a relative rate method**:** $Y_R = (X_J \times X_K)$, where $X_J$ stands for the measured rate constant ratio ($K_{TEST}/K_{REF}$) and $X_K$ for the experimental value of "$K_{REF}$" [5]. Clearly, the result $y_R$ is obtained as**:** $(y_R \pm \varepsilon_R) = (x_J \pm u_J) \times (x_K \pm u_K)$. Even, the simultaneous determination (cf. Eq. **1c**) of above said "$Y_{Li}$" and "$Y_B$" as "$Li_2BO_2^+$" requires [8,9] only **two** different measured data, i.e.**:** $(y_{Li} \pm \varepsilon_{Li}^S) = f_{Li}^S([x_J \pm u_J], [x_K \pm u_K])$; **and** $(y_B \pm \varepsilon_B^S) = f_B^S([x_J \pm u_J], [x_K \pm u_K])$, (with, say**:** ***J*** as "**55/57**" and ***K*** as "**56/57**"). —— However suppose that the measurement-uncertainties are, *irrespective* of the systems [5,8,9], fixed**:** $u_J = u_K = 0.2\%$. Then, we enquire, should also all system-specific [5,8,9] output-uncertainties be 0.2% (at least, is**:** $\varepsilon_R = \varepsilon^S_{Li} = \varepsilon^S_B$)? Moreover, how should $\varepsilon^S_{Li}$ and $\varepsilon^S_B$ be evaluated? Do we have a, instead of system-specific [1] or intensive [16,17] method as the Monte Carlo, simpler means for evaluating **any** "$\varepsilon_m$" (viz.**:** $\varepsilon_R$, $\varepsilon^S_{Li}$, $\varepsilon^S_B$, etc)?

In short, say**:** $\boldsymbol{Y_A = f_A(X_J)}$, $\boldsymbol{Y_B = f_B(X_J)}$, $\boldsymbol{Y_C = f_C(X_J)}$, etc. Then should (for a given *estimate* $\boldsymbol{x_J}$) the errors in the estimates as**:** $\boldsymbol{y_A = f_A(x_J)}$, $\boldsymbol{y_B = f_B(x_J)}$, $\boldsymbol{y_C = f_C(x_J)}$, vary from one another? Clearly, the errors could be varying **provided** the different *theoretical* means of measurements (i.e. the SSRs: $\boldsymbol{f_A}$, $\boldsymbol{f_B}$, $\boldsymbol{f_C}$) are, like different *physical* $X_J$ measurement techniques, differently sensitive. The theoretical techniques as **Eq. 1** are therefore studied here for their possible properties. We thus present the $\varepsilon_m$ versus $u_i(s)$ relationship (section 2), validate it, and/ or show (sections 3-3.4) how it helps authenticate *any* result (evaluate a question as above). Even, we clarify from an experimental viewpoint whether the standard value ($\rho_m$) of the output-uncertainty can differ from its actual value ($\varepsilon_m$), i.e. what might govern $\varepsilon_m$ (sections 3.1-3.2), and provide further insight into the features of evaluations (section 3.3).

It should however be noted that a given SSR is sometimes for the convenience of discussion referred to below by alone "$Y_m$", viz. "$Y_G = f_G(X_J,X_K,X_L)$" as "$Y_G$".

## 2. FORMALISM: THE $\varepsilon_m$ and $u_i(s)$ RELATIONSHIP

It is well-known [11] that the error in $y_m$ (cf. Eq. 1) can even numerically differ from the error in a corresponding $x_i$. Thus, e.g. the absolute error $\delta Y_m$ in an $y_m$ obtained by the Eq. 1b above could be accounted for as [11]:

$$\delta Y_m^{Theo} = \sum_{i=1}^{N} \left(\frac{\partial Y_m}{\partial X_i}\right) \Delta X_i, \text{ (for a given } m) \tag{3}$$

where $\Delta X_i$ stands for any kind of errors whatever in $x_i$. That is, "$\delta Y_m$" cannot vary for whether**:** (i) "$\Delta X_i$" is by origin either systematic or statistical or both, (ii) the errors "$\Delta X_1$, $\Delta X_2$, .. $\Delta X_N$" are inter-correlated, etc. However, as the true index of an error is its relative value, we rewrite Eq. 3:

$$\delta_m^{Theo} = \left(\frac{\delta Y_m^{Theo}}{Y_m}\right) = \sum_{i=1}^{N}\left[\left\{\left(\frac{\partial Y_m}{\partial X_i}\right)\left(\frac{X_i}{Y_m}\right)\right\}\left(\frac{\Delta X_i}{X_i}\right)\right] = \sum_{i=1}^{N} M_i^m \, \Delta_i \qquad (4)$$

Or, we may define [8] the error-ratio "$|\delta_m^{Theo}| \big/ \sum_{i=1}^{N} |\Delta_i|$" as the collective error multiplication factor ($C_m^{Theo}$), and more usefully express Eq. 4 as:

$$|\delta_m^{Theo}| = \left|\sum_{i=1}^{N} M_i^m \, \Delta_i\right| = \sum_{i=1}^{N} |M_i^m| \, |\Delta_i| = C_m^{Theo} \sum_{i=1}^{N} |\Delta_i| \qquad (5)$$

where the individual error multipliers $\{M_i^m\}$are defined as:

$$M_i^m = \left(\frac{\partial Y_m}{\partial X_i}\right)\left(\frac{X_i}{Y_m}\right), \quad i = 1, 2, ..., N \text{ (for a given } m) \qquad (6)$$

It may be pointed out that Eqs. 3-5, even though introduced in relation to the evaluations as Eq. 1b, represent all types of cases. Thus consider, e.g. an Eq. 1c**:** $f_i(Y_A, Y_B, Y_C) = X_i$ ($i$ = 1, 2, 3). Then the $\delta_m^{Theo}$-formulae ($m$ = $A$, $B$ and $C$), which were derived elsewhere [8] via the process of solving a set of differential equations, could be seen having exactly the same form as Eq. 4. Of course Eq. 3-5 will, for any simple derived system as Eq. 1a, also simplify, viz.:

$$|\delta_m^{Theo}| = \left|\sum_{i=1}^{N} M_i^m \, \Delta_i\right| = |M_i^m \, \Delta_i| = C_m^{Theo} \, |\Delta_i|, \quad (i = m = N = 1) \qquad (5a)$$

Eq. 5a clarifies that the translation of even a single measured data into any *derived* result (i.e.**:** $x_i \rightarrow y_m$, cf. Eq. 1a$^{/}$) is accomplished by the transformation of the error $\Delta_i$ (if any, in $x_i$) into the error $\delta_m$ through a multiplier ($M_i^m$), however. That is, for measurement-accuracy alone to be the yardstick, the result $y_m$ will be subject to **over** or **under** estimation.

Nevertheless, Eq. 6 defines "$M_i^m(s)$" to be the theoretical constant(s) for a given SSR, thereby making the corresponding error multiplication factor ($C_m$) to be even a priori predicted:

$$C_m^{Theo} = \frac{|\delta_m^{Theo}|}{\sum_{i=1}^{N} |\Delta_i|} = \frac{|\sum_{i=1}^{N} M_i^m \Delta_i|}{\sum_{i=1}^{N} |\Delta_i|} = {\left|M_1^m + \sum_{i=2}^{N} M_i^m (\Delta_i/\Delta_1)\right|}\Big/{1 + \sum_{i=2}^{N} |\Delta_i/\Delta_1|} \quad (7)$$

Naturally, Eq. 7 will for SSRs represented by Eq. 1a reduce:

$$C_m^{Theo} = \frac{|\sum_{i=1}^{N} M_i^m \Delta_i|}{\sum_{i=1}^{N} |\Delta_i|} = \frac{|M_i^m \Delta_i|}{|\Delta_i|} = |M_i^m|, \ (i = m = N = 1) \quad (7a)$$

Further, if: $|\Delta_i| = |\Delta_i^{Max}| = u_i$, then: $|\delta_m| = |\delta_m^{Max}| = \varepsilon_m$. That is, Eq. 5 may be rewritten as:

$$\varepsilon_m = C_m^{Theo} \sum_{i=1}^{N} u_i = \sum_{i=1}^{N} |M_i^m| \, u_i, \quad \text{(for a given } m\text{)} \quad (8)$$

Or, for: $u_i = u_1$ (with: $i = 2, 3, \ldots N$):

$$\varepsilon_m = C_m^{Theo} N u_i = (\sum_{i=1}^{N} |M_i^m|) \, u_i, \quad \text{(for a given } m\text{)} \quad (9)$$

It may be pointed out that, in Eq. 3, all higher order factors (viz.: $\frac{\partial^P Y_m}{\partial X_i^P} \Delta X_i^P$ , with $P \geq 2$) are neglected. However, as for a *linear* SSR: $\frac{\partial^2 Y_m}{\partial X_1^2} = 0$, $\frac{\partial^2 Y_m}{\partial X_1 \partial X_2} = 0$, etc., the actual error ($\delta_m$) and the uncertainty ($\varepsilon_m$) in any corresponding result ($y_m$) should exactly be accounted for by Eq. 4 and Eq. 8, respectively. Further, 'minimization of error ($\Delta_i$)' is the general experimental motto, i.e. it is expected that: $(\Delta_i)^P \cong 0$. Therefore, results by even non-linear SSRs should also be explicable by the theory (Eqs. 3-9). In fact, that the output-uncertainty is represented by Eq. 8 rather than by Eq. 2 was indicated previously (cf. section 5.2.2 in [1]). Yet, we should cross-check our findings here.

## 3. RESULTS AND DISCUSSION: VALIDATION OF THE THEORY (Eq. 8/9)

Any real world [2-10] result $y_m$ is shaped through a theoretical task as Eq. 1. Therefore the process, for verifying whether really the characteristics of $y_m$ vary with Eq. 1, should clearly be theoretical. However, the data $x_i(s)$ are required (cf. Eqs. 3-9) to correspond $X_i$-standard(s). Of course, for specific systems [2,4,8], such data are also available. Yet, it is here believed to be worth examining the implications of Eqs. 3-9 using at least certain *flawless* data, viz. those (simulated for arbitrary but general possible derived systems represented by the $X_i$-*standards* as: $X_J^T$ = 10.0, $X_K^T$ = 5.0, and $X_L^T$ = 77.5, and the *constants* as: α = 10.13 and β = 5.8) in Table 1.

### 3.1 Direct measurement: uncertainty ($u_i$)

Suppose that all the data in Table 1 are obtained by a single method of measurement. Then, it may be pointed out that the nature of the data gives no indication of bias in the measurements. That is to say that the imprecision of the implied method of measurement, though so high as 0.01%, is the sole cause for inaccuracy ($u_i$). In other words, the error ($\Delta_i$) in a relevant unknown estimate $x_i$ is to be considered ±0.01%. Further, the table clarifies that the error of repetitive measurements (e.g. $\Delta_J^T$) is not the same as the corresponding standard deviation ($\sigma_J^T$).

### 3.2 Indirect measurement: distinctions between parameters as $u_i$, $\varepsilon_m$ and $\rho_m$

Table 2 presents, for each of the SSRs described therein (and substituting $X_i$ by the $X_i^T$ above, and hence $x_i$ by $x_i^T$ from Table 1), the evaluated $Y_m$-*values*, and their parameters as the probable error $\rho_m$, the actual error $\delta_m$, the error multiplication factor $C_m$, etc. That is, whether the features of an $y_m$ can ever be different from those of its $x_i(s)$ is illustrated in Table 2. The table shows that: $|\delta_m| \neq |\Delta_i|$ (with $i = J$ or, if applicable, $K$, or $L$, cf. $Y_1$-$Y_9$), thereby clarifying that the output-uncertainty $\varepsilon_m$ cannot generally be represented by the measurement-uncertainty $u_i$.

Again, the measurement-errors (cf. Table 1) are by origin random, i.e.: $u_i = \sigma_i$. However, the output-error "$|\delta_m|$" has exceeded at least in some cases "$\rho_m$" (cf. $Y_1$-$Y_4$ in Table 2), which imply that the uncertainty $\varepsilon_m$ cannot also in general be represented by its standard value ($\rho_m$).

### 3.2.1 $\varepsilon_m$ vs. $Y_m$: is the number ($N$) of $X_i$-variables (measurements) a key factor?

It is shown (cf. Eq. 6) above how, for given an Eq. 1 (SSR), the rates ($\{M_i^m\}$)) of variations of $Y_m$ as a function of $\{X_i\}$ could really a priori be predicted. That is, it is already indicated above why can the output-error $\delta_m$ vary with *alone* the functional nature of $Y_m$ (comparison between the $\{Y_m\}$ as: $Y_1$-$Y_6$ or: $Y_8$-$Y_{10}$ for a given example no. in Table 2). Further, for each of the SSRs in Table 2, the $M_i^m$- and $\varepsilon_m$-values are furnished in Table 3. Therefore, whether or not the results in Table 2 are explicable by the theory can easily be examined. Thus a result (e.g.: $\delta_4$ = **0.012**, cf. example no. 1) for $Y_4$, or for any other ***linear*** case, can be seen to be ***exactly*** accountable by Eq. 4 (as: $\delta_4^{Theo} = (M_J^4\,\Delta_J + M_K^4\,\Delta_K) = (2 \times 0.008) + (-1 \times 0.004) = 0.012$). Even the results obtained by a sensitive *non-linear* SSR as $Y_5$ (with: $M_J^5$ = **103.6**) could be well accounted for. Thus e.g. the variation between: $\delta_5$ = **0.839** (example no. 1) and its predicted value: $\delta_5^{Theo}$ = **0.834** is small and explicable in terms of the neglected factors as "$(\Delta_i)^P$, with: $P \geq 2$" in Eq. 4.

Moreover, in terms of uncertainty (Eq. 8/ 9), any result in Table 2 should be accountable as: $|\delta_m| \leq \varepsilon_m$. For example, Table 1 implies the *method-specific* measurement-uncertainty ($u_i$) to be **0.01%**, i.e. (cf. Table 3): $\varepsilon_4$ = 0.03% (whereas: $\delta_4$ = 0.012%); and: $\varepsilon_5$ = 1.05% (though: $\delta_5$ = 0.839%, cf. above). And, for $u_i$ to be "*example-specific*-$|\Delta_i^{Max}|$", viz. **0.008%** (cf. example no. 1 in Table 1), $\varepsilon_4$ = **0.024%** and $\varepsilon_5$ = **0.84%** (i.e. even then: $\delta_4 < \varepsilon_4$, and: $\delta_5 = \varepsilon_5$).

In any case, it is in Table 3 demonstrated that, and also clarified (in terms of the governing factor(s), $M_i^m$*(s)*) why, the uncertainty $\varepsilon_m$ varies with alone the function "$f_m$" (i.e. why: $\frac{\varepsilon_m^{Highest}}{\varepsilon_m^{Lowest}} = {}^{\varepsilon_5}/_{\varepsilon_3} = 104.95$, cf. the cases as: $Y_m = f_m(X_J,X_K)$) or simply for the operator (${}^{\varepsilon_4}/_{\varepsilon_3} = 3$, cf. $Y_1$-$Y_4$). In other words, the $\varepsilon_m$ is shown to be decided by the (***description*** of the) SSR rather than by the (number "$N$" of) measurements. Even the results ($y_8$-$y_{10}$ in Table 2) for the measurement-systems as: $Y_m = f_m(X_J)$, i.e. which reflect the SSR "$Y_8$" as a fixed error **source**, "$Y_9$" as an error **sink**, and "$Y_{10}$" to be a **non-interfering** agent (but which are reciprocated by the respective uncertainties $\boldsymbol{\varepsilon_8}$, $\boldsymbol{\varepsilon_9}$ and $\boldsymbol{\varepsilon_{10}}$ in Table 3), are in corroboration of the said statement. It may over and above be noted that (cf. Table 2): $\boldsymbol{Y_6 = f_6(X_J, X_K)}$, and: $\boldsymbol{Y_7 = f_7(X_J, X_K, X_L)}$. However (see Table 3): $\varepsilon_6 = \mathbf{8.25}u_i$, whereas: $\varepsilon_7 = \mathbf{1.48}u_i$. That this is the fact can be verified as follows. Let: $u_i = 0.01\%$, so that: $\varepsilon_6 = 0.0825\%$, and: $\varepsilon_7 = 0.0148\%$. However, say: (**1**) $\Delta_J = \Delta_K = 0.01\%$ and $\Delta_L = -0.01\%$; (**2**) $\Delta_J = \Delta_K = -0.01\%$ and $\Delta_L = 0.01\%$; (**3**) $\Delta_J = \Delta_K = \Delta_L = 0.01\%$; (**4**) $\Delta_J = \Delta_K = \Delta_L = -0.01\%$, etc. Then, clearly, the net input error is higher for "$y_7$" ($\sum_{i=J}^{L} |\Delta_i| = \mathbf{0.03}\%$) than for "$y_6$" ($\sum_{i=J}^{K} |\Delta_i| = \mathbf{0.02}\%$). However, one can verify that: $|\boldsymbol{\delta_7}| < |\boldsymbol{\delta_6}|$, viz. (for case nos. **1** and **2**): $|\boldsymbol{\delta_7}| = 0.0148\%$, but: $|\boldsymbol{\delta_6}| \cong 0.0825\%$. Or while: $|\boldsymbol{\delta_7}| = 0.01\%$, $|\boldsymbol{\delta_6}| \cong 0.0825\%$ (cf. case nos. **3** and **4**).

### 3.2.2 $Y_m$-families and $\varepsilon_m$: is the $Y_R$-system [5] or the Boyle's Law [18] represented by "$Y_1$"?

As clarified in Table 3, "$\boldsymbol{M_i^m}$" can turn out either *sensitive* to system-defining $\boldsymbol{X_i}$*-value* (i.e. strictly SSR-specific), or *ever fixed* (i.e.: $|\boldsymbol{M_i^m}| = \mathbf{1}$). Thus, say, an SSR, which is characterized by "$|\boldsymbol{M_i^m}| = \mathbf{1}$, $(\boldsymbol{i} = \mathbf{1, 2} \ldots \boldsymbol{N})$" belongs to the family no.: F.1; and an SSR, for which *any* "$|\boldsymbol{M_i^m}| \neq \mathbf{1}$", is a member of the family no.: F.2. Then, for alone F.1, Eq. 8/ 9 reduces to: $\boldsymbol{\varepsilon_m} = \sum_{i=1}^{N} u_i = \boldsymbol{f_m(\{u_i\})}$. Thus, if only the SSR (i.e. irrespective of whatever might the desired $Y_m$ and the measured $X_i(s)$ stand for) is given, it should be known beforehand whether the output-uncertainty will be *fixed* (as **F.1**) by the $u_i(s)$ only, or *vary* ($\boldsymbol{\varepsilon_m} = \sum_{i=1}^{N} |M_i^m|\, u_i = \boldsymbol{f_m(\{X_i, u_i\})}$, cf. **F.2**) with even the $X_i$*-value(s)*. That is, proper a priori planning of experiments could then be

possible. Anyway, the SSRs: $Y_1$, $Y_2$ and $Y_{10}$ are F.1 members, but all the other $Y_m$-systems in Table 2/ 3 belong to F.2.

In fact, the significance of family-features could be better understood in terms of the above mentioned case [5] of determining the OH-acetone reaction rate constant ($Y_R$), i.e. one for which the data on $X_i$ standards are difficult to be obtained. —— The SSR [5]: $Y_R = (X_J \times X_K)$ is, by nature, no different from the SSR: $Y_1 = (X_J \times X_K)$ in Table 2. That is, one can verify that: $M_J^R = M_J^1 = 1$ and: $M_K^R = M_K^1 = 1$(cf. Table 3). In other words, for given $\{u_i\}$, the result $y_R$ should be as uncertain as $y_1$ (cf. Eq. 8): $\boldsymbol{\varepsilon_R} = \boldsymbol{\varepsilon_1} = (|M_J^1|u_J + |M_K^1|u_K) = (u_J + u_K)$, irrespective of whatever might be the values of corresponding[5] $X_J$ and $X_K$. Of course, the veracity of our prediction could also be judged using reported estimates, viz. (cf. Eq. 4 and Table 1 **in [5]** for 303°K): $\boldsymbol{x_J} = (5.23 \pm 0.54) = (5.23 \pm 10.3\%)$ and $\boldsymbol{x_K} = (3.983\times10^{-14} \pm 20\%)$; and: $\boldsymbol{y_R} = ([2.08 \pm 0.22]\times10^{-13}) = (2.08\times10^{-13} \pm 10.3\%)$. Clearly the $x_J$ and $x_K$ were acquired [5] by different means with uncertainties as high as **10.3**% and **20**%, respectively. And, the desired result $y_R$ was there reported (cf. Table 1 **in [5]**) against the "$u_J$" **alone**. However, that (cf. above): $\varepsilon_R = (u_J + u_K) = \mathbf{30.3}\%$ can be verified (on: $X_J = \mathbf{5.23}$ and $X_K = \mathbf{3.983\times10^{-14}}$, and hence on: $Y_R = (X_J \times X_K) = \mathbf{2.08\times10^{-13}}$) as follows:

**1.** $y_R = (x_J \times x_K) = ([X_J + \Delta_J^{Max}] \times [X_K + \Delta_K^{Max}]) = ([X_J + \mathbf{0.103}X_J] \times [X_K + \mathbf{0.20}X_K])$

$= (5.77 \times [4.78\times10^{-14}]) = 2.76\times10^{-13} = (Y_R + \mathbf{0.32}Y_R)$;

**2.** $y_R = ([X_J - \mathbf{10.3\%}] \times [X_K - \mathbf{20\%}]) = (4.69 \times [3.187\times10^{-14}]) = 1.49\times10^{-13} = (Y_R - \mathbf{28\%})$;

**3.** (*All* other error-combinations (with: $|\Delta_J^{Max}| = \mathbf{10.3\%}$ and: $|\Delta_K^{Max}| = \mathbf{20\%}$)" imply: $|\boldsymbol{\delta_R}| \le \mathbf{30.3\%}$, viz.): $y_R = ([X_J + \mathbf{0.103}X_J] \times [X_K - \mathbf{0.20}X_K]) = 1.84\times10^{-13} = (Y_R - \mathbf{11.6\%})$.

Similarly, using any other data-set [5], it could be demonstrated that: $\varepsilon_R = (u_J + u_K)$. However, the $u_K$ was so high as 20%, and the $u_J$ was reported to vary (with the measurement-temperature in the range) as: **4.70-19.2%** (**cf.** Table 1 **in [5]**). Therefore: $\varepsilon_R$ = **24.70-39.2%**, i.e. only more accurate data than those in [5] should help to better unfold the reaction mechanism there. In any case, it should be clear that "$\boldsymbol{\varepsilon_m = (u_J + u_K)}$" holds for ***any*** system as "$\boldsymbol{Y_m = (X_J \times X_K)}$". Yet, it may be worth elaborating further on the issue in terms of gas-laws (see APPENDIX 2, where on, Eq. 8/9 is applied for both *random* and *systematic* $u_i$-sources).

### 3.2.3 Limiting $C_m$ and/ or $\delta_m$: classifications of indirect measurement systems (SSRs)

As shown in Table 1, **no** $\boldsymbol{x_i}$ is absolutely accurate (i.e.: $\boldsymbol{x_i^T \neq X_i^T}$). Yet, as shown for certain cases in Table 2, "$\boldsymbol{y_m = Y_m}$" (e.g.: $\boldsymbol{\delta_1 = 0}$, cf. the example no. 3 for $Y_1$). Then, are those cases wrongly presented? —— Actually, it is already clarified above (cf. Eq. 5) that, if somehow the error multiplication factor $C_m$ turns out to be zero, the output-error $\delta_m$ will equal to zero. And this should be true, even though Eq. 7a predicts $C_m$ to be an SSR-specific non-zero constant. That is, it is also a fact that *no* **Eq. 1a** can lead: $(\boldsymbol{\Delta_i \neq 0}) \rightarrow (\boldsymbol{\delta_m = 0})$, cf. Eq. 5a. For example (cf. $Y_9$ in Table 3): $C_9^{Theo} = |M_J^9|$ = **0.073**, and (cf. the example **no. 5** in Table 1): $|\Delta_J|$ = **0.0001**. Therefore: $|\delta_9^{\mathrm{Theo}}| = 0.073\ |\Delta_J|$ = $\mathbf{7.3x10^{-6}}$. This is why the error $\delta_9$ is, though from the practical viewpoint zero, not shown as zero in Table 2.

However, as Eq. 7 implies, $C_m$ is a constant (either **zero**, or >**0**) of experimental error-ratio(s). That is *any* **Eq. 1b/ 1c** can cause: $\boldsymbol{\{\Delta_i \neq 0\} \rightarrow (\delta_m = 0)}$. And the corresponding requirement, for the systems e.g. as "$Y_m = f_m(X_J, X_K)$" in Table 2, is (cf. Eq. 4):

$$\frac{\Delta_J}{\Delta_K} = -\frac{M_K^m}{M_J^m} \tag{10}$$

Eq. 10 explains why, for all the experiments (say, corresponding to $Y_1$), the $C_1$, and hence the $\delta_1$, did not turn out zero. In fact, that any such system (***N* = 2**) has got a singular possibility for the $C_m$ to be zero is better clarified in Fig. 1, which describes the predicted variations of $\{C_m\}$ corresponding to "$Y_m = f_m(X_J, X_K)$" and also "$Y_m = f_m(X_J)$" in Table 2, and hence which helps validate all those results there (compare the observed $C_m$-values in Table 2 with their predicted values). It may however be recalled that the $C_8$, $C_9$ and $C_{10}$ are *independent* of $\Delta_J$ (cf. Eq. 7a). Yet, if**:** $\Delta_J = 0$, then**:** $\delta_m = (C_m \times \Delta_J) = 0$ (cf. Eq. 5a and, $Y_8$-$Y_{10}$). Which is why the $C_8$ or $C_9$ or $C_{10}$ (cf. the inserts in Fig. 2) is at "$\Delta_J = 0$" projected as zero. In any case, for**:** ***N* = 1**, the $C_m$ can **never** be zero. And, for**:** ***N* = 2**, the $C_m$ can but only under **a given condition** equal zero. Then should the possibility "$C_m = 0$" be, in a case as $Y_7$ (***N* = 3**) in Table 2, just twice? Interestingly the chances are, as dictated by Eq. 10a below and exemplified in Fig 2, **innumerable**:

$$\frac{\Delta_J}{\Delta_K} = -\left(\frac{M_K^m}{M_J^m} + \frac{M_L^m}{M_J^m}\frac{\Delta_L}{\Delta_K}\right) \tag{10a}$$

Fig. 2 depicts $C_7$ as a function of the error-ratios, $\Delta_J/\Delta_K$ and $\Delta_L/\Delta_K$. Clearly, any XZ plane (defined by a given $\Delta_L/\Delta_K$) describes the variation**:** $C_7$ vs. $\Delta_J/\Delta_K$; and an YZ plane (identified by a fixed $\Delta_J/\Delta_K$) depicts**:** $C_7$ vs. $\Delta_L/\Delta_K$. Further, it is important noting that the point $\Delta_i/\Delta_K = 0$ (with either**:** $i = J$, or**:** $i = L$, cf. Fig. 1/ 2) does not denote**:** $\Delta_i = \Delta_K = 0$, but it refers to**:** $\Delta_i = 0$ and $\Delta_K$ as any non-zero number. However, Fig. 2 clarifies that every XZ (or YZ) plane has got within or outside the figure-dimension a discrete point as "$C_7 = 0$". Yet, why didn't $C_7$ corresponding to any of the five different sets of observations in Table 2 equal zero is readily explicable. Say**:** $\Delta_L/\Delta_K = -0.6452$ (cf. example no. 1 in Table 2). Then, Eq. 10a yields**:** $\Delta_J/\Delta_K = -5.5$. However, Table 2 shows**:** $\Delta_J/\Delta_K = 2.0$, thereby explaining why the corresponding $C_7$ is non-zero.

Now, it may be recollected that any two SSRs should also in terms of their parameter(s) as "$M_i^m$" (cf. Eq. 6) be distinguishable from one another. Again, Eq. **1a** and Eq. **1b** are shown here above to be different by really class-property. Thus say, all possible SSRs of one independent variable (Eq. **1a**) constitute a group**:** Gr. (I). Then, all SSRs with more than one experimental variable, and/ or represented by Eq. **1b** and Eq. **1c** as well, should also fit into another single group (Gr. (II)). This is because the $C_m$ corresponding to any $Y_m$ but represented by **Eq. 1c** had already been established [8] to be a constant (either **zero**, or >**0**) for, like a case of Eq. 1b (cf. Fig. 1 or 2), given experimental error-ratios ($\Delta_J$**:** $\Delta_K$**:** $\Delta_L$**: …**) only.

### 3.2.4 Specific aspects of Gr. (I) and Gr. (II): $C_m$ and $\varepsilon_m$ values

In the case of Gr. (I), the $C_m$ is predicted (Eq. 7a) and also verified above (cf. $Y_8$-$Y_{10}$ in Table 2) to be an SSR-specific constant. Therefore, for Gr. (I), Eq. 8/ 9 might be rewritten as:

$$\varepsilon_m^{\mathrm{I}} = C_m^{Theo}\, N\, u_i = C_m^{Theo}\, u_i = |M_i^m|\, u_i \qquad (9a)$$

where the superscript "I" refers to the Gr. (I).

However, the Gr. (II) $C_m$ is shown to vary with experimental errors (cf. Fig. 1/ 2 and Table 2 for**:** $Y_1$-$Y_7$). Nevertheless the highest value ($C_m^{Max}$) that it can take is, as clarified by Fig. 1/ 2, prefixed as the SSR-specific-**highest**-"$|M_i^m|$" (which is, henceforth, denoted by**:** ${}^H|M_i^m|$), viz.**:** $C_1^{Max} = {}^H|M_J^1| = {}^H|M_K^1| = 1.0$, $C_4^{Max} = |M_J^4| = 2.0$, etc. cf. Table 3. However can we, like the case of Gr. (I), express the Gr. (II)-uncertainty ($\varepsilon_m^{\mathrm{II}}$) as below?

$$\varepsilon_m^{\mathrm{II}} = C_m^{Theo}\, N\, u_i = {}^H|M_i^m|\, N\, u_i \qquad (9b)$$

Clearly, for**:** $|M_1^m| = |M_i^m|$, $i$ = 2, 3,… $N$ (i.e. for any **F.1** family member of Gr. (II), e.g. $Y_1$ or $Y_2$ in Table 2), Eq. 9 and Eq. 9b are equivalent. However, for **F.2** members (viz. $Y_3$-$Y_7$), Eq. 9b offers exaggerated estimates. Consider, e.g. the SSR: $Y_4$. Then, as clarified in Table 3: $\varepsilon_4 = 3u_i$, but: $\varepsilon_4^{II} = 4u_i$. Even, as discussed below, the ***true*** $\varepsilon_m$ (Eq. 8/ 9) could be **mistaken** as higher.

It is above exemplified (e.g. for $Y_1$, $Y_6$ and $Y_7$, cf. sections 3.2.1- 3.2.2) that, if really ***all*** the different measurement errors ($\Delta_1$, $\Delta_2$, …, $\Delta_N$) should equal **by magnitude** to their highest possible values ($u_1$, $u_2$, …, $u_N$, respectively) and also simultaneously turn out to be parallel **by sign** to their respective multipliers ($M_1^m$, $M_2^m$, …, $M_N^m$), then and only then the actual error ($\delta_m$) in the result ($y_m$) will be equaling its highest possible value ($\varepsilon_m$). Such an occurrence, though may stand as trivial, cannot be ruled out. Therefore, the Gr. (II) $\varepsilon_m$ could but only be ***risked*** to believe having a value somewhat *less* than its *true* value as the Eq. 8/ 9.

### 3.3 Evaluation of $\varepsilon_m$ in practice: choice of experimental conditions and/ or variables (*$X_i$'s*)

It may first be emphasized that, though the purpose is to evaluate an *unknown* as $Y_m$, the SSR (Eq. 1) cannot be unknown. Therefore, the required knowledge of SSR-specific parameter(s) ($M_i^m(s)$, cf. Eq. 6) can always be acquired in terms of real or theoretical $X_i$-standards (cf. sections 3.1-3.2). That is, Eq. 8/ 9 could be used to predict a priori the $u_i(s)$ required for achieving a preset accuracy ($\varepsilon_m$) in the desired result ($y_m$). For illustration, say that the desired $Y_m$-system is by its features similar to the SSR: $Y_2$ in Table 2/ 3, and the result $y_m$ is required to be as accurate as ***p***%. Then, as: $\varepsilon_2 = 2u_i$, the measurement-accuracy is needed to be at least two fold better ($u_i \leq$ ***0.5p***). Or, if the measurement-procedure(s) and thus the achievable $u_i(s)$ should be prefixed, then the uncertainty $\varepsilon_m$ can also really be predetermined (cf. Eq. 8/ 9), i.e. the result $y_m$ can at least be correctly validated as: $|\boldsymbol{\delta}_m| \leq \boldsymbol{\varepsilon}_m$ (with $\boldsymbol{\delta}_m$ as the unknown error in $y_m$). In any case, accuracy of an $y_m$ can be crosschecked by evaluating $M_i^m(s)$ on actually measured data $x_i(s)$, thereby inferring whether any more planned experimentation is necessary.

Further, how the $X_i$ variable(s) and/ or the SSR may among different possible alternatives in a case [2,8-10] be judiciously selected is indicated by the systems as $Y_6$ (with: $N$ = **2**, and $\varepsilon_6$ = **8.25**$u_i$) and $Y_7$ (with: $N$ = **3**, and $\varepsilon_7$ = **1.48**$u_i$) in Table 2. It may thus be noted that: $Y_6 = Y_7$. That is, by the SSRs: $Y_6$ and $Y_7$, it is meant two different methods for determining however a single parameter of a given system. Therefore, if the additional measurement ($X_L$) should pose no problem, then the preferred process of evaluation is represented by the SSR: $Y_7$.

However, an $Y_m$-system of the type as SSR "$Y_6$" should be worth elaborating. As: $\varepsilon_6 = 8.25u_i$, the measurement-accuracy is required to be ≈**10** times better than that to be desired for the result $y_m$. However, as the curve $C_6$ in Fig. 1 indicates, the pre-evaluation of measurement-conditions on standards can help improve the accuracy $\varepsilon_6$. For example, "$(\Delta_J/\Delta_K) \leq (-3.0)$" yields: $C_6 \leq$ **1.0**, which should in turn give: $\boldsymbol{\varepsilon_6 \leq 2u_i}$ (cf. Eq. 9). At least, it might not be impossible achieving: $\boldsymbol{\varepsilon_6 \approx 4u_i}$ (as, for either: $\boldsymbol{(\Delta_J/\Delta_K) \geq 5.0}$, or: $\boldsymbol{(\Delta_J/\Delta_K) \leq (-1.5)}$, $C_6 \leq 2.0$).

#### 3.3.1 Requirement for an evaluation to be successful vs. that for a chemical reaction to be spontaneous (ΔG < 0): a highlight

By success, it is here meant that: $\varepsilon_m \leq u_i$. Thus an SSR, which implies "$\sum_{i=1}^{N} |M_i^m|$" to be ≤ 1 (cf. Eq. 9), can a priori be guaranteed to lead the evaluation to success. Again, it is well-known that any exothermic reaction ($\boldsymbol{\Delta H < 0}$) is by nature spontaneous: $\boldsymbol{(\Delta G = \Delta H - T\Delta S) < 0}$. That is to say that a successful evaluation and an exothermic reaction might, by characteristics, be considered as parallel. If so, then an **undesirable** SSR ($\boldsymbol{\sum_{i=1}^{N}|M_i^m| > 1}$) should be said parallel to an **endothermic** reaction ($\boldsymbol{\Delta H > 0}$). Clearly, in the latter case, the reaction will take place provided the temperature ($T$) is raised so high that: $T\Delta S > \Delta H$. Similarly, here, the controlling factor is the method-sensitive measurement-uncertainty "$u_i$", which should if at all be feasible ensured so small that it yields acceptable "$\varepsilon_m$" by overriding "$\sum_{i=1}^{N}|M_i^m| > 1$".

Further, a specified product of an endothermic reaction might sometime be obtainable by an alternative exothermic path ($\Delta H < 0$). Similarly the measured variable(s) $X_i(s)$ and/ or the SSR should in a possible case [2,8-10] be so chosen that $\varepsilon_m$ is $\leq u_i$, at least, the ratio "$\varepsilon_m/u_i$" is lower than that offered by any alternative process (cf. SSRs $Y_6$ and $Y_7$).

### 3.4 The uncertainty $\varepsilon_m$ and typical real world evaluations

#### 3.4.1 Gr. (I) cases with and without a possible choice of the working-variable $X_i$

It is clarified above that accuracy ($\varepsilon_m$) of determining an $Y_m$ is really preset by the nature (i.e.: $M_i^m$, cf. Eq. 8/ 9) of the corresponding SSR (here: $\boldsymbol{Y_m = f_m(X_i)}$). Therefore, we will here elaborate on only "$M_i^m$" (cf. Eq. 6) of interested cases, viz. for [2] evaluating $^6$Li/$^7$Li abundance ratio ($Y_{Li}$) from measured abundance ratio ($X_i$) of a pair ($i$) of isotopic $Li_2BO_2^+$ ions**:** $Y_{Li} = f_{Li}(X_i)$. However, irrespective of the isotopic $Li_2BO_2^+$ pair "***i***", the function [2] "$f_{Li}$" could be shown to relate (like "$f_8$" or "$f_9$" in Table 2) the **F.2** family. That is the rate-of-variation ($M_i^{Li}$) of $Y_{Li}$ with $X_i$, and/ or the uncertainty of evaluation ($\boldsymbol{\varepsilon_{Li} = |M_i^{Li}|u_i}$, cf. Eq. 9a), will depend on $|X_i|$, i.e. really on the constituent elemental isotopic abundances (**CEIAs**). Thus, for illustration, consider all constituents (Li, B and O) to be *natural*. Then the "$Li_2BO_2^+$" mass spectrum could be shown [2], even theoretically [19], to project m/z (56, 57) and m/z (55, 57) as the most and the second most abundant ion-pairs, respectively. Thus, say [2]**:** $\boldsymbol{X_{56/57}}$ = 0.413533 and**:** $\boldsymbol{X_{55/57}}$ = 0.04805, which could in turn be shown to mean that**:** $M_{56/57}^{Li}$ = **2.5**; and**:** $M_{55/57}^{Li}$ = **0.9**, respectively. That is, if the measurement procedure is so established that**:** $u_{55/57} = u_{56/57}$, then the desired result ($y_{Li}$) will turn out better accurate (than even the measured data $x_i$) for using m/z (55, 57) rather than m/z (56, 57) as the monitor-pair ($\boldsymbol{i}$). In fact [2]: $\boldsymbol{Y_{Li}} = f_{Li}(X_{56/57}) = 0.0832$ or: $\boldsymbol{Y_{Li}} = f_{Li}(X_{55/57}) = 0.0832$, but e.g. "$\boldsymbol{x_{56/57}} = (\boldsymbol{X_{56/57}} + \mathbf{0.1\%}) = 0.413947$" gives: $y_{Li} = (\boldsymbol{Y_{Li}} + \mathbf{0.25\%}) = 0.0834$; and "$\boldsymbol{x_{55/57}} = (\boldsymbol{X_{55/57}} + \mathbf{0.1\%}) = 0.048098$" yields: $y_{Li} = (\boldsymbol{Y_{Li}} + \mathbf{0.09\%}) = 0.083273$. —— And, even for ***independently*** determining the constituent $^{10}B/^{11}B$ ratio $Y_B$ (i.e. for: $\boldsymbol{Y_B = f_B(X_i)}$), the m/z (55, 57) is predicted as the better monitor-pair, i.e.: $M^B_{55/57}$ = **1.2**, whereas: $M^B_{56/57}$ = **1.6**. Further, that our predictions are facts **(**i.e. that an 'a priori' analysis of SSR-specific property(s) can really help in either properly designing the required experiments or, correctly validating the desired result**)** could be verified in terms of experimental data [2] on standards.

We now apply our uncertainty consideration to an apparently involved case (SSR) as the correlation [3] of the second virial coefficient ($Y_W$) of water with temperature ($X_T$ °K): $\boldsymbol{Y_W = Y_0 \sum_{n=1}^{4} a_n (X_T/100)^{b_n}}$ (with $Y_0$, $a_n$, and $b_n$ as constants). We consider this correlation as the perfect one, and inquire whether the uncertainty ($u_T$) in monitoring $X_T$ should exactly be the uncertainty ($\varepsilon_W$) in the predicted value ($y_W$) of $Y_W$. —— It could be shown that the present SSR also relate to the **F.2** family, i.e. the rate ($M_T^W$) of $Y_W$ vs. $X_T$ variation will itself be dictated by the system-temperature ($X_T$). For example, $M_T^W$ takes (for $X_T$ = 275, 300, 325, 2500, 3000, 3500 °K) the values as –5.61, –5.0, –4.48, 1.27, 0.75, 0.49, respectively. Then (cf. Eq. 9a): $\varepsilon_W = 5.61u_{275°K}$ (or $\varepsilon_W = 4.48u_{325°K}$), which implies that $y_W$ should be at a larger error than the error actually incurred in measuring a **lower** temperature. Therefore, this might be the basic reason why an experimental $Y_W$*-value* had deviated [3] from its predicted value at a lower temperature. Further the harmony, recorded [3] between experimental and predicted $Y_W$*-values* at any relatively **higher** temperature, is also echoed in our findings here: $\varepsilon_W = 1.27u_{2500°K}$, $\varepsilon_W = 0.49u_{3500°K}$, etc.

Clearly, even such observations assert that the measurement-accuracy ($u_i$) alone cannot be the basis for validating a derived result.

**3.4.2 Gr. (II) systems: why may $y_m$ vary [7] with even alone the evaluation model?**

We again consider the case of isotopic analysis as $\mathrm{Li_2BO_2^+}$ (cf. section 3.4.1), but assume the purpose to be the *simultaneous* determination [8a,9] of $Y_{Li}$ and $Y_B$ by employing m/z (55, 57) and m/z (56, 57) as the monitor ion-pairs (J and K, respectively, cf. **Eq. 1c**). The corresponding SSRs (i.e. "$\boldsymbol{f_J(Y_{Li},Y_B) = X_J}$" and "$\boldsymbol{f_K(Y_{Li},Y_B) = X_K}$"), it could be shown, belong to the **F.2** family. And, their parameters take (for all *natural* constituents, i.e. for: $X_J = X_{55/57}$ = **0.048050** and: $X_K = X_{56/57}$ = **0.413533**, cf. above) the values as: (i) ${}_{Li,B}^{\;\;S}M_J^{Li}$ = 1.8 and ${}_{Li,B}^{\;\;S}M_K^{Li}$= -2.5; and (ii) ${}_{Li,B}^{\;\;S}M_J^{B}$= -1.2 and ${}_{Li,B}^{\;\;S}M_K^{B}$= 3.4 (cf. Eq. 6); where the prefix as either "*S*" or "*Li,B*" is meant for distinguishing the present case from the above said individual evaluations [2] of $Y_{Li}$ and $Y_B$.

Now, say: $u_J = u_K = u_i$. Then, the uncertainties of determination are predicted to be (cf. Eq. 9): $\varepsilon_{\mathrm{Li}}^{\mathrm{S}} = (\sum_{i=J}^{K} |{}_{Li,B}^{\;\;S}M_i^{Li}|)u_i$ = **4.3**$u_i$ and $\varepsilon_{\mathrm{B}}^{\mathrm{S}} = (\sum_{i=J}^{K} |{}_{Li,B}^{\;\;S}M_i^{B}|)u_i$ = **4.6**$u_i$. Further, like the above case [2], it could be shown that: **(i)** for "$X_J$ = **0.048050** and, $X_K$ = **0.413533**", the solutions of the set of SSRs conform to the true values ($Y_{Li}$ = **0.0832** and, $Y_B$ = **0.2473**); but **(ii)** for measured estimates, e.g. "$\boldsymbol{x_J}$ = ($\boldsymbol{X_J}$ **– 0.1%**) = 0.048002, and $\boldsymbol{x_K}$ = ($\boldsymbol{X_K}$ **+ 0.1%**) = 0.413947**"** one obtains: $\boldsymbol{y_{Li}}$ = ($\boldsymbol{Y_{Li}}$ **– 0.43%**) = 0.082846, and $\boldsymbol{y_B}$ = ($\boldsymbol{Y_B}$ **+ 0.46%**) = 0.248422. Over and above, the facts that: **(1)** $\varepsilon_{\mathrm{Li}}^{\mathrm{S}}$ differs from either "$\varepsilon_B^S$" or "$\varepsilon_{Li}$ **(**of Gr. (I)**)**"; and **(2)** $\varepsilon_{\mathrm{B}}^{\mathrm{S}}$ varies from $\varepsilon_B$ (cf. Gr. (I): $\boldsymbol{Y_B = f_B(X_i)}$, *i* = ***J*** or ***K***), confirm that $\varepsilon_m$ can vary for the SSR *alone*.

Here, it may also be of interest to enquire whether the predictions above vary with **CEIAs**. Thus, suppose that only lithium is enriched to 95.6% in $^6$Li (i.e.: $Y_{Li}$ **= 21.73**, and $Y_B$ **= 0.2473**). Then, Eq. 9 gives: $\varepsilon_{\mathrm{Li}}^{\mathrm{S}}/\mathrm{u_i}$ = **3** and $\varepsilon_{\mathrm{B}}^{\mathrm{S}}/\mathrm{u_i}$ = **290**, i.e. even negligible errors in the measured data ($x_J$ and $x_K$) are predicted to cause the results, specifically $y_B$, to be useless. Thus, it could be shown e.g. that "$\mathbf{x_J}$ = ($X_J$ + **0.1%**), and $\boldsymbol{x_K}$ = ($X_K$ + **0.1%**)" yield**:** $\boldsymbol{y_{Li}}$ **=** ($Y_{Li}$ – 0.2%) = 21.6894, but $\boldsymbol{y_B}$ **=** ($Y_B$ **+ 29%**) = 0.32. Similarly, "$\mathbf{x_J}$ = ($X_J$ + **0.1%**), and $\boldsymbol{x_K}$ = ($X_K$ – **0.1%)**" give**:** $\boldsymbol{y_{Li}}$ **=** ($Y_{Li}$ **+ 0.3%**) = 21.7911, and, $\boldsymbol{y_B}$ = ($Y_B$ – ***11.5%***) = *0.219*. In fact, such a real world evaluation was reported [8a] to yield a relatively accurate estimate for $Y_{Li}$ and an absurd $y_B$. It may thus be emphasized that, if a derived result should be judged by acceptable measurement-uncertainty ($u_i$) ***alone***, then the "$y_B$" means attributing a very ***odd*** value to "$Y_B$".

Anyway, for ensuring $y_B$ to be accurate (say) as**:** $\boldsymbol{\varepsilon_B^S}$ = ***p%***, the measurements are needed to be so accurate that**:** $u_i$ **≤ 0.0033*p*%**. That is to say that such an experiment **(**here precisely using m/z (55, 57) and m/z (56, 57) as the required ***monitor ion-pairs*)** should, even when there could be no alternative for, be worth abandoning. Fortunately, it can in this (enriched Li) case be shown [8a] that m/z (55, 57) and m/z (56, 57) do not, from even alone measurement-viewpoint, conform as the preferred monitor pairs (J and K). The most abundant ions-pairs, and hence the desirable monitor pairs, are [8a] m/z (54, 55) and m/z (55, 56). Furthermore, m/z (54, 55) and m/z (55, 56) are predicted ($\varepsilon_{Li}^S$ = **1.04**$u_i$ and $\varepsilon_B^S$ = **1.05**$u_i$), and can also in experimental [8a] terms be shown, to yield the results ($y_{Li}$ and $y_B$) as accurate as the measured data ($x_{54/55}$ and $x_{55/56}$).

Further, by the number (***N***) of $X_i$-variables, the Eq. **1c** [ 8a,9], and the Eq. **1b** (as $\boldsymbol{Y_1}$ in Table 2 and the rate constant [5] $\boldsymbol{Y_R}$), systems are comparable. However, while $\varepsilon_{Li}^S$ and $\varepsilon_B^S$ will depend on $X_i$*-values*, $\varepsilon_1$ or $\varepsilon_R$ will remain ever fixed as**:** $\varepsilon_1 = \varepsilon_R$ = **2$\boldsymbol{u_i}$**. This supplements the finding above that "$\varepsilon_m$" is governed by SSR(s) rather than by measurements. —— In fact, the case [6] of determining critical micelle concentration ($Y_C$) and the corresponding standard free energy of micellization ($Y_G$) should better illustrate the point here. The evaluations [6] could be represented as (**Eq. 1b**): $Y_C = f_C(X_J,X_K,X_L)$, and: $Y_G = f(Y_C) = f_G(X_J,X_K,X_L)$. That is, *one and the same set* of experimental data were required [6] for determining both $Y_C$ and $Y_G$. Yet, it could be shown that: $\varepsilon_\mathbf{C} \neq \varepsilon_\mathbf{G}$. Actually, both "$f_C$" and "$f_G$" relate (like, e.g. "$f_7$" in Table 2) to the **F.2** family, i.e.: $\varepsilon_C = (\sum_{i=J}^{L} |M_i^C|)u_i = \boldsymbol{a}u_i$; and: $\varepsilon_G = (\sum_{i=J}^{L} |M_i^G|)u_i = \boldsymbol{b}u_i$; with ***a*** and ***b*** have **(**for the different sets of estimates "$x_J$, $x_K$, and $x_L$" presented in [6]**)** the values as **2.0-4.8** and **0.23-0.77**, respectively. Thus, if: $u_J = u_K = u_L$, then the estimate $y_C$ is more (2-5 times) ***inaccurate***, and the corresponding $y_G$ is however more (1.3-4 times) **accurate**, than any of their measured data ($x_J$, $x_K$, and $x_L$ [6]). However [6]: $u_J \neq u_K \neq u_L$, e.g. while: $u_J$ = **6.74**% (i.e.: $\boldsymbol{x_J}$ = 0.89 ± 0.06); $u_K$ = **0.98**% ($\boldsymbol{x_K}$ = 102 ± 1) and: $u_L$ = **4.29**% ($\boldsymbol{x_L}$ = 21.0 ± 0.9). Then, as Eq. 8 predicts: $\varepsilon_C$ = **9.1**% (i.e.: $\boldsymbol{y_C}$ = [0.011 ± 0.001]), and: $\varepsilon_G$ = **1.3**% (i.e.: $\boldsymbol{y_G}$ = –[17.12 ± 0.22]). Even, that these are facts could be verified like above cases, viz.: [$\boldsymbol{x_J} = (\boldsymbol{x_J}$ **+ 6.74%**), $\boldsymbol{x_K} = (\boldsymbol{x_K}$ **– 0.98%**) and $\boldsymbol{x_L} = (\boldsymbol{x_L}$ **+ 4.29%**)] yield: $\boldsymbol{y_C} = (\boldsymbol{y_C}$ **+ 9.3%**) = 0.012; and: $\boldsymbol{y_G} = (\boldsymbol{y_G}$ **– 1.3%**) = –16.9. In fact, $X_J$, $X_K$ and $X_L$, are *inter-correlated* [6]. Thus, it is also confirmed that "$\boldsymbol{\varepsilon_m}$" is independent of the *nature* of, however, $\boldsymbol{X_i}$*-variables*.

Even our comparative findings (in terms of $Y_C$ and $Y_G$ here, and/ or $Y_6$ and $Y_7$ in Table 2) can help clarify, it may be pointed out, why the results for a given derived variable ($Y_m$) but which were evaluated [7] by employing different data evaluation models (i.e. by using different values for the required *constants* there) varied from one another.

Finally, for why the accounting of a derived result is ***in exact terms of its SSR*** significant, it may be mentioned that "$|M_i^m|$" corresponding to one or more real world SSRs was reported [8b,8c] a number so small as $\approx 10^{-5}$ or as large as $\approx 3\text{x}10^4$. That is to note that the nature of an SSR could be so governing that the result will, even for significant measurement error(s) $u_i(s)$, turn out ≈100% accurate (viz. $\varepsilon_m \approx \mathbf{10^{-5}} \boldsymbol{Nu_i}$, cf. Eq. 9b). Or, the result $y_m$ may, for really a negligible level of error $u_i$ in its $x_i(s)$, misrepresent (as, $\varepsilon_m \approx \mathbf{10^{4}} \boldsymbol{Nu_i}$) the variable ($Y_m$) it stood for.

## 4. CONCLUSIONS

The above study clarifies that no SSR ($X_i(s) \rightarrow Y_m(s)$) can without checking its property be considered to behave as a perfect tool for transforming the measured data, $x_i(s)$, into the desired result(s), $y_m(s)$. In support, it is demonstrated that $y_m$ can depending simply on the nature of the SSR turn out less, or even more, reliable than $x_i(s)$. That is, as the purity of a chemical product is (for given purities of reactants) decided by the reaction(s) involved in production, the uncertainty $\varepsilon_m$ in the estimate $y_m$ is (for given uncertainty(s) $u_i(s)$ in $x_i(s)$) shown to be dictated by the SSR(s) shaping the $y_m$. Moreover, a given chemical reaction can by alone its properties be distinguished from some other. Similarly, why may at all $\varepsilon_m$ vary as a function of the theoretical tool as SSR is explained by identifying a given SSR with a given (set of) parameter(s), $M_i^m(s)$, which preset(s) the relative rate(s) of variation(s) of $Y_m$ with $X_i(s)$. Again, as all chemical reactions fall for a given feature under different categories, the study above led us to place all the SSRs of one and more than one experimental variable ($X_i$) into two groups, Gr. (I) and Gr. (II), respectively. The identifying parameter is the ratio ($C_m$) of the *error-in-result* to the *net-input-error*. In case of Gr. (I), $C_m$ is demonstrated to be an SSR-specific **theoretical** constant (>**0**). However, corresponding to any Gr. (II) $Y_m$, the $C_m$ is pointed out to be a constant (either **zero** or >**0**) for a given set of **experimental** data ($\{x_i\}$) only.

Further, any evaluation can be described to involve only two different steps**:** the measurement (of $X_i(s)$) and the result-shaping ($X_i(s) \rightarrow Y_m(s)$). However, the (latter) theoretical task is really by its role and effect inseparable from any individual experimental step. For example, the general requirement for a measurement to be carried out is to a priori look into the pros and cons of every discrete task it consisted of (viz. sample preparation, choice of instrumental settings, actual measurement of $X_i$'s, etc.), thereby making the task to yield to its purpose. Yet, an experimental-step may either leave the already accumulated error unchanged, or add to it, or reduce it, or even nullify the same to define the overall experimental error ($\Delta_i$). Likewise, any valid (set of) SSR(s) could always be seen to yield $y_m^{True}(s)$ from $x_i^{True}(s)$. However, by a measured data ($x_i$), it is meant that: $x_i = (x_i^{True} + \Delta_i) = (X_i + \Delta_i)$. Therefore, the result ($y_m$) should also signify that: $y_m = (y_m^{True} + \delta_m) = (Y_m + \delta_m)$. Thus it is demonstrated above that the error-shaping, $\Delta_i(s) \rightarrow \delta_m(s)$, is an integral part of the corresponding (SSR-dictated) process of result-shaping: $x_i(s) \rightarrow y_m(s)$. Further, a required experimental step could be bracketed with its purpose (effects). Similarly, the study above shows that the SSR can a priori be marked as either a non-modifier of the input-error ($\Delta_i$), or a sort of additional error-source, or even an error ($\Delta_i$) sink, which will eventually suggest measures to be taken in designing the required experiment(s) and/ or the evaluation itself. Essentially, any **desired** data translation, $(x_1, x_2, \dots x_N) \rightarrow y_m$, is signified above as a **given** uncertainty transformation, $(u_1, u_2, \dots u_N) \rightarrow \varepsilon_m$:

$$\varepsilon_m = \sum_{i=1}^{N} |M_i^m|\, u_i$$

If all different $\{X_i\}$ are measured by a single technique of uncertainty $u_i$, then:

$$\varepsilon_m = \left(\sum_{i=1}^{N} |M_i^m|\right) u_i$$

It is also exemplified above that, "$x_i(s) \rightarrow y_m$" and "$u_i(s) \rightarrow \varepsilon_m$", stand for complementary and desired (SSR regulated) systematic changes. If "$N$" is unity, one obtains: $\varepsilon_m = |M_i^m|\, u_i = C_m u_i$; i.e. the **Gr. (I)** behavior (viz. why the collective error multiplication factor $C_m$ should also be a SSR-specific theoretical constant) is explained for. Moreover, Gr. (I) makes it easy to understand why the required data-accuracy $u_i$ (for achieving a desired accuracy, $\varepsilon_m$, in the result $y_m$) is really preset by the SSR involved. It is also clarified above why, for $N \geq 2$ (**Gr. II**), $C_m$ is controlled by the errors-in-data. Actually the $C_m$ varies, for *alone* a variation in error-ratios ($\Delta_1$**:** $\Delta_2$**:** $\Delta_3$**: …**) from one experiment to another (but irrespective of whether the total error**:** $\sum_{i=1}^{N} |\Delta_i|$ varies), within a range as**: zero** $\leq C_m \leq {}^{H}|M_i^m|$, (with "${}^{H}|M_i^m|$" as the value of $Y_m$-specific "highest-$M_i^m$"). Clearly, "$C_m = 0$" implies that *any* Gr. (II) SSR can cause**:** $\{\Delta_i \neq 0\}_{i=1}^{N} \rightarrow (\delta_m = 0)$, i.e. can lead "$\{x_i\}$" but for **a given pattern of their errors** to yield the "$y_m^{True}$". However, whether the error "$|\delta_m|$" can ever exceed the net experimental error "$\sum_{i=1}^{N} |\Delta_i|$", and/ or to what extent, is dictated by the specified $Y_m$ (as**:** $C_m^{Max} = {}^{H}|M_i^m|$). This in turn explains why, even for given data $\{x_i\}$ and/ or their uncertainties $\{u_i\}$, the resultant-uncertainty ($\varepsilon_m$) varies from one $Y_m$ to another. It is further shown that $\varepsilon_m$ cannot vary for whether the data ($\{x_i\}$, and thus $\{u_i\}$) are inter-correlated [6], and/ or even if $\{u_i\}$ involve bias-contribution [18].

However, irrespective of whether a given system (SSR) should belong to Gr. (I) or Gr. (II), it is outlined above how to judiciously choose the experimental conditions and/ or (if applicable) the monitor-variable(s), and hence to make the evaluation a success. It is also clarified how, in a case where preplanning of experiments should be difficult, the assessment of the system-specific $\varepsilon_m$ (i.e. validation of desired result $y_m$) is to be made up theoretically, viz. by incorporating some error(s) in the measured data $x_i(s)$ and observing the corresponding rate(s) of variation(s) in the desired result $y_m$.

Over and above, our study should help incorporate in any relevant data evaluation model the provision for correctly ascertaining the uncertainty ($\varepsilon_m$) in the desired result ($y_m$), such as against the uncertainty(s) $u_i(s)$ in the measured data $x_i(s)$. Of course, the provision means feeding of also SSR-specific $M_i^m$-formula(s). However the reason, why could $y_m$ vary with alone the evaluation model [7], will then be clear (cf.**:** $Y_6$ and $Y_7$, and/ or**:** $Y_C$ and $Y_G$, above). It may in fact be suggested that, given the SSR, one can first check whether the same belongs to the **F.1** family (i.e. whether**:** $\{|\boldsymbol{M_i^m}| = \mathbf{1}\}_{\boldsymbol{i=1}}^{\boldsymbol{N}}$, so that**:** $\boldsymbol{\varepsilon}_{\mathbf{m}} = \sum_{\boldsymbol{i}\,=\,\mathbf{1}}^{\boldsymbol{N}} \boldsymbol{u_i}$) or to the **F.2** family (i.e. whether *any* single "$M_i^m$" is decided by $X_i(s)$, and hence when**:** $\boldsymbol{\varepsilon}_{\mathbf{m}} = \sum_{\boldsymbol{i}\,=\,\mathbf{1}}^{\boldsymbol{N}} |\boldsymbol{M_i^m}|\, \boldsymbol{u_i}$). Clearly, for **F.1**, the feeding of $M_i^m$-formula(s) is not required. A representative example of "**F.1**" is the Boyle's (ideal gas) system, and that of the "**F.2**" is the van der Waals system.

## ACKNOWLEDGEMENT

The author sincerely thanks Dr. S. K. Das and Dr. A. K. Dhara for constructive suggestions.

**Table 1**

Supposedly measured values for $X_J$, $X_K$ and $X_L$ standards ($X_J^T$, $X_K^T$, and $X_L^T$, respectively): examples distinguishing between measurement precision ($\sigma_i^T$), error ($\Delta_i^T$) and uncertainty (achievable accuracy, $u_i$)

| Ex. No. | Mean $x_J^T$ ± $\sigma_J^T$(%) (% Error: $\Delta_J^T$) | Mean $x_K^T$ ± $\sigma_K^T$(%) (% Error: $\Delta_K^T$) | Mean $x_L^T$ ± $\sigma_L^T$(%) (% Error: $\Delta_L^T$) | Reflected expt. (example)-specific uncertainty: ±$u_i$ |
|---|---|---|---|---|
| 1 | 10.0008 ±0.009 (0.008) | 5.0002 ±0.006 (0.004) | 77.498 ±0.005 (-0.0026) | $\pm\sigma_J^T$ |
| 2 | 9.9996 ±0.007 (-0.004) | 4.9996 ±0.01 (-0.008) | 77.5008 ±0.004 (0.0010) | $\pm\sigma_K^T$ |
| 3 | 10.0006 ±0.006 (0.006) | 4.9997 ±0.008 (-0.006) | 77.5025 ±0.005 (0.0032) | $\pm\sigma_K^T$ |
| 4 | 10.00029 ±0.0033 (0.0029) | 4.99998 ±0.008 (-0.0004) | 77.4969 ±0.007 (-0.004) | $\pm\sigma_K^T$ |
| 5 | 9.99999 ±0.0006 (-0.0001) | 5.0005 ±0.01 (0.01) | 77.4985 ±0.0035 (-0.0019) | $\pm\sigma_K^T$ |

**Table 2**

Different derived variables $\{Y_m\}$ **(**and their estimates**:** $\{y_m\}$, statistical errors**:** $\{\rho_m\}$, actual errors**:** $\{\delta_m\}$ and error multiplication factors**:** $\{C_m\}$**)** but corresponding to certain given measured variables ($X_J$, $X_K$ and $X_L$, cf. Table 1) and constants ($\alpha$ = 10.13 and $\beta$ = 5.8)

| $Y_m$-formula (Eq. 1) and its true value | Ex. No. | i) $\Sigma_i\lvert\Delta_i\rvert$<br>ii) $\Delta_J/\Delta_K$<br>iii) $\Delta_L/\Delta_K$ | $y_m$<br>$\pm\rho_m$ (%) | i) $\delta_m$ (%)<br>ii) $C_m$ |
|---|---|---|---|---|
| $Y_1 = f_1(X_J, X_K)$<br>$= X_J \times X_K$<br>= 50.0 | 1 | i) 0.012<br>ii) 2.0 | 50.006<br>±0.0108 | i) 0.0120003<br>ii) 1.000027 |
| | 2 | i) 0.012<br>ii) 0.5 | 49.994<br>±0.0122 | i) -0.0120<br>ii) 0.999973 |
| | 3 | i) 0.012<br>ii) –1.0 | 50.0<br>±0.010 | i) 0.0<br>ii) 0.0 |
| | 4 | i) 0.0033<br>ii) –7.25 | 50.00125<br>±0.00865 | i) 0.0025<br>ii) 0.757572 |
| | 5 | i) 0.0101<br>ii) –0.01 | 50.00495<br>±0.01002 | i) 0.0099<br>ii) 0.980197 |
| $Y_2 = f_2(X_J, X_K)$<br>$= \frac{X_J}{X_K}$<br>= 2.0 | 1 | i) 0.012<br>ii) 2.0 | 2.00008<br>±0.0108 | i) 0.0040<br>ii) 0.333320 |
| | 2 | i) 0.012<br>ii) 0.5 | 2.00008<br>±0.0122 | i) 0.004<br>ii) 0.333360 |
| | 3 | i) 0.012<br>ii) –1.0 | 2.00024<br>±0.010 | i) 0.0120<br>ii) 1.000060 |
| | 4 | i) 0.0033<br>ii) –7.25 | 2.000066<br>±0.00865 | i) 0.0033<br>ii) 1.000004 |
| | 5 | i) 0.0101<br>ii) –0.01 | 1.999798<br>±0.01002 | i) -0.0101<br>ii) 0.99990 |
| $Y_3 = f_3(X_J, X_K)$<br>$= X_J + X_K$<br>= 15.0 | 1 | i) 0.012<br>ii) 2.0 | 15.001<br>±0.006325 | i) 0.006667<br>ii) 0.555556 |
| | 2 | i) 0.012<br>ii) 0.5 | 14.9992<br>±0.00573 | i) -0.00533<br>ii) 0.444444 |
| | 3 | i) 0.012<br>ii) –1.0 | 15.0003<br>±0.0048 | i) 0.0020<br>ii) 0.166667 |
| | 4 | i) 0.0033<br>ii) –7.25 | 15.00027<br>±0.00346 | i) 0.0018<br>ii) 0.545455 |
| | 5 | i) 0.0101<br>ii) –0.01 | 15.00049<br>±0.003357 | i) 0.003267<br>ii) 0.323432 |
| $Y_4 = f_4(X_J, X_K)$<br>$= X_J - X_K$<br>= 5.0 | 1 | i) 0.012<br>ii) 2.0 | 5.0006<br>±0.019 | i) 0.012<br>ii) 1.0 |
| | 2 | i) 0.012<br>ii) 0.5 | 5.0<br>±0.0172 | i) 0.0<br>ii) 0.0 |
| | 3 | i) 0.012<br>ii) –1.0 | 5.0009<br>±0.0144 | i) 0.018<br>ii) 1.50 |
| | 4 | i) 0.0033<br>ii) –7.25 | 5.00031<br>±0.0104 | i) 0.0062<br>ii) 1.878788 |
| | 5 | i) 0.0101<br>ii) –0.01 | 4.99949<br>±0.01007 | i) -0.0102<br>ii) 1.009901 |

**Table 2** continued

| $Y_m$-formula (Eq. 1) and its true value | Ex. No. | i) $\Sigma_i \lvert \Delta_i \rvert$ ii) $\Delta_J/\Delta_K$ iii) $\Delta_L/\Delta_K$ | $y_m$ $\pm\rho_m$ (%) | i) $\delta_m$ (%) ii) $C_m$ |
|---|---|---|---|---|
| $Y_5 = f_5(X_J, X_K)$ $= \frac{X_K}{\alpha - X_J} - X_J$ = 28.461538 | 1 | i) 0.012 ii) 2.0 | 28.70 ±0.94 | i) 0.839 ii) 69.95 |
| | 2 | i) 0.012 ii) 0.5 | 28.34 ±0.72 | i) -0.42 ii) 35.32 |
| | 3 | i) 0.012 ii) –1.0 | 28.64 ±0.624 | i) 0.616 ii) 51.36 |
| | 4 | i) 0.0033 ii) –7.25 | 28.547 ±0.343 | i) 0.30 ii) 91.1 |
| | 5 | i) 0.0101 ii) –0.01 | 28.46244 ±0.064 | i) 0.003153 ii) 0.312218 |
| $Y_6 = f_6(X_J, X_K)$ $= \frac{X_J X_K}{\beta - X_K}$ = 62.5 | 1 | i) 0.012 ii) 2.0 | 62.523 ±0.044 | i) 0.037 ii) 3.084 |
| | 2 | i) 0.012 ii) 0.5 | 62.461 ±0.073 | i) -0.062 ii) 5.164 |
| | 3 | i) 0.012 ii) –1.0 | 62.477 ±0.0583 | i) -0.037 ii) 3.124 |
| | 4 | i) 0.0033 ii) –7.25 | 62.50 ±0.0581 | i) 0.0 ii) 0.0 |
| | 5 | i) 0.0101 ii) –0.01 | 62.545 ±0.073 | i) 0.072 ii) 7.173 |
| $Y_7 = f_7(X_J, X_K, X_L)$ $= X_L - X_J - X_K$ = 62.5 | 1 | i) 0.014581 ii) 2.0 iii) -0.6452 | 62.497 ±0.0.0064 | i) –0.0048 ii) 0.3292 |
| | 2 | i) 0.013032 ii) 0.5 iii) -0.1290 | 62.5016 ±0.0051 | i) 0.00256 ii) 0.1964 |
| | 3 | i) 0.015226 ii) –1.0 iii) -0.5376 | 62.5022 ±0.0063 | i) 0.00352 ii) 0.2312 |
| | 4 | i) 0.0073 ii) –7.25 iii) 10.0 | 62.49663 ±0.0087 | i) -0.0054 ii) 0.7386 |
| | 5 | i) 0.012035 ii) –0.01 iii)-0.1935 | 62.49801 ±0.0044 | i) –0.0032 ii) 0.2646 |
| $Y_8 = f_8(X_J) = \frac{\alpha - X_J}{\beta}$ = 0.022414 | 1 | i) 0.008 | 0.02228 ±0.70 | i) -0.6154 ii) 76.923 |
| | 5 | i) 0.0001 | 0.022416 ±0.046 | i) 0.007692 ii) 76.923 |
| $Y_9 = f_9(X_J)$ $= \sqrt{\frac{X_J}{\alpha} + \beta}$ = 2.60521915 | 1 | i) 0.008 | 2.605234 ±0.00065 | i) 0.000582 ii) 0.072723 |
| | 5 | i) 0.0001 | 2.605219 ±0.000044 | i)-0.000007 ii) 0.072723 |
| $Y_{10} = f_{10}(X_J)$ $= (\alpha - \beta)X_J$ = 43.30 | 1 | i) 0.008 | 43.3035 ±0.009 | i) 0.008 ii) 1.0 |
| | 5 | i) 0.0001 | 43.299957 ±0.0006 | i) -0.0001 ii) 1.0 |

**Table 3**

Characteristic theoretical constants $\{M_i^m\}$ and the predicted $\{\varepsilon_m\}$ corresponding to all the different $\{Y_m\}$, i.e. the SSRs, in Table 2

| $Y_m$ | $M_J^m$ | $M_K^m$ | $\varepsilon_m$ (Eq. 9) |
|---|---|---|---|
| $Y_1$ | $M_J^1 = 1.0$ | $M_K^1 = 1.0$ | $\varepsilon_1 = 2u_i$ |
| $Y_2$ | $M_J^2 = 1.0$ | $M_K^2 = -1.0$ | $\varepsilon_2 = 2u_i$ |
| $Y_3$ | $M_J^3 = \frac{X_J}{X_J + X_K} = 0.6667$ | $M_K^3 = \frac{X_K}{X_J + X_K} = 0.3333$ | $\varepsilon_3 = u_i$ |
| $Y_4$ | $M_J^4 = \frac{X_J}{X_J - X_K} = 2.0$ | $M_K^4 = \frac{X_K}{X_J - X_K} = -1.0$ | $\varepsilon_4 = 3u_i$ |
| $Y_5$ | $M_J^5 = \frac{X_J\left[X_K - (\alpha - X_J)^2\right]}{(\alpha - X_J)[X_K - (\alpha - X_J)X_J]}$ $= 103.598753$ | $M_K^5 = \frac{X_K}{X_K - (\alpha - X_J)X_J}$ $= 1.351351$ | $\varepsilon_5 = 104.95u_i$ |
| $Y_6$ | $M_J^6 = 1.0$ | $M_K^6 = \frac{\beta}{\beta - X_K} = 7.25$ | $\varepsilon_6 = 8.25u_i$ |
| $Y_7$ * | $M_J^7 = \frac{X_J}{X_J + X_K - X_L}$ $= -0.16$ | $M_K^7 = \frac{X_K}{X_J + X_K - X_L}$ $= -0.08$ | $\varepsilon_7 = 1.48u_i$ |
| $Y_8$ | $M_J^8 = \frac{X_J}{X_J - \alpha} = -76.923077$ | - | $\varepsilon_8 = 76.923u_i$ |
| $Y_9$ | $M_J^9 = \frac{X_J}{2(\alpha\beta + X_J)} = 0.072723$ | - | $\varepsilon_9 = 0.073u_i$ |
| $Y_{10}$ | $M_J^{10} = 1.0$ | - | $\varepsilon_{10} = u_i$ |

*: $M_L^7 = \frac{X_L}{X_L - X_J - X_K} = 1.24$.

## APPENDIX 1: Notations

Input (*measured*/ independent) and output (*desired*/ dependent) variables are rather by norm denoted here differently, viz. as: $X_i$ and $Y_m$ (and their estimates as: $x_i$ and $y_m$), respectively. Again, an evaluation might involve more than one measured variable, and also in some cases enable the simultaneous determination of several output-variables. Thus, both input and output variables are at the outset subscripted. For example, by "$X_i$, with: $i = J$ and $K$" (or "$Y_B$ and $Y_{Li}$"; or "$\{X_i\}_{i=1}^{2}$"; or: "$Y_m$, with: $m$ = 1 and 2"; or so), it is referred two different variables.

Similarly, for clarity, any specific *input* and *output parameters* are here distinguished by even notations. Thus, $\boldsymbol{\Delta X_i}$ and $\boldsymbol{\delta Y_m}$ refer to the (true) *absolute errors*; $\boldsymbol{\Delta_\mathrm{i}}$ and $\boldsymbol{\delta_\mathrm{m}}$ to the **relative errors**; $\boldsymbol{u_i}$ and $\boldsymbol{\varepsilon_\mathrm{m}}$ (i.e.: "$|\Delta_\mathrm{i}^\mathrm{Max}|$" and "$|\boldsymbol{\delta}_\mathrm{m}^\mathrm{Max}|$") to the ***relative uncertainties***; $\boldsymbol{\sigma_\mathrm{i}}$ and $\rho_\mathrm{m}$ to the relative scatters (*relative standard/ probable errors*); **…** in the estimates: $x_i$ and $y_m$ (of the input and output variables: $X_i$ and $Y_m$), respectively. For example, the (relative) error, the uncertainty and the scatter in the estimate $\boldsymbol{y_1}$ (of an output variable $\boldsymbol{Y_1}$) are referred to here as: $\boldsymbol{\delta_1}$, $\boldsymbol{\varepsilon_1}$, and $\rho_1$, respectively. Likewise, *limiting* and *predicted values*, of the error "$\delta_\mathrm{m}$" are denoted as: "$\delta_\mathrm{m}^\mathrm{Lim.}$" and "$\delta_m^{Theo}$", respectively. It may also here be pointed out that (even for an established method of $X_i$-measurement), the true error $\Delta_\mathrm{i}$ is likely to vary from one experiment to another. However the corresponding highest possible *value* ($|\Delta_\mathrm{i}^\mathrm{Max}|$, i.e. "$u_i$"), is expected to be unique, really. Thus, for any desired result $y_m$, the corresponding uncertainty $\varepsilon_\mathrm{m}$ should (though take a value different from "$u_i$", cf. the text) also accordingly be fixed.

Further, for: $Y_m = f_m(\{X_i\}_{i=1}^{N})$, the rate-of-variation of $Y_m$ as a function of $X_i$ is referred as "$M_i^m$" (e.g. $M_{55/57}^{Li}$, where the SSR is: $Y_{Li} = f_{Li}(X_{55/57})$).

## APPENDIX 2: THE SSR "$Y_m = (X_J \times X_K)$" AND THE GAS LAWS

Let $X_J$ be the pressure and $X_K$ the volume of *one mole* of ideal gas at $T$ °K. Then, according to the Boyle's Law [18], the product "$(X_J \times X_K)$" is a constant (say, $Y_T$) equaling "$RT$" (with "$R$" as the Gas constant). That is the immediate implication of the Boyle's Law is that, for given the gas-pressure $X_J$, the volume $X_K$ should be known, and the vice-versa. Say**:** $T$ = 273.16°K **(**i.e.**:** $Y_T = RT$ = (0.08205447 × 273.16) = 22.414 lit.-atm.**)**, and**:** $X_J$ = 400 atm. Then, it is expected that**:** $X_K = (Y_T / X_J)$ = (22.414 / 400) = 0.056035 liter. Further, "$\mathbf{Y_T = (X_J \times X_K)}$" could like the SSR "$\mathbf{Y_1}$" be shown to imply**:** $\boldsymbol{M_J^T = M_K^T = 1}$ (cf. Eq.6). In other words, "$x_J = (X_J \pm u_J)$ and $x_K = (X_K \pm u_K)$" should yield**:** $y_T = (Y_T \pm \boldsymbol{\varepsilon_T}) = (Y_T \pm [\boldsymbol{u_J} + \boldsymbol{u_K}])$. Thus, e.g. "$x_J = (X_J$ + **0.1%**) = 400.4 and $x_K = (X_K$ + **0.1%**) = 0.056091" give**:** $y_T = (x_J \times x_K)$ = 22.459 = ($Y_T$ + **0.2%**).

However, the experimental verification of the Boyle's Law is difficult [18]. For example, volume ($x_J^{N_2}$) of 1 mole of nitrogen gas, at 273.16°K and under the preset pressure ($x_K^{N_2}$) of 400 atm., was measured [18] to be 0.0703 liter. Therefore**:** $y_T^{N_2} = (x_J^{N_2} \times x_K^{N_2})$ = 28.12 lit.-atm., i.e. the error ($\delta_T^{N_2} = ([y_T^{N_2}/Y_T] - 1)$ = **25.5%**) is too high to be accounted for by the possible *random* errors in the estimates ($x_J^{N_2}$ and, $x_K^{N_2}$). However, why should $\delta_T^{N_2}$ be unimaginably high?

Any real gas is, unlike the ideal gas, characterized by species-specific coulomb forces [18]. Thus, neither "$x_J^{N_2}$" can stand for the ideal gas pressure "$X_J$", nor "$x_K^{N_2}$" for the volume "$X_K$". But, at best (i.e. for *random* errors in measurements to be **zero**)**:** $x_J^{N_2} = (X_J - p^{N_2})$ and $x_K^{N_2} = (X_K + v^{N_2})$; with "$p^{N_2}$" and "$v^{N_2}$" as the *systematic* errors in $x_J^{N_2}$ and $x_K^{N_2}$, respectively. Therefore, **for a mole of real gas**, the Boyles Law could be re-expressed as:

$$(X_P + p)\,(X_V - v) = RT = Y_T \qquad \text{(A.1)}$$

where $X_P$ and $X_V$ are the **observable** (i.e. real gas) pressure and volume, and $p$ and $v$ are their deviations from the ideal gas pressure and volume $X_J$ and $X_K$, respectively, at $T$ °K.

However, first, imagine "*p*" and "*v*" to be *fixed instrumental biases,* so that the measured responses (say, $r_P$ and $r_V$) should be corrected to yield**:** $\boldsymbol{x_P = (r_P + p) = x_J}$ and $\boldsymbol{x_V = (r_V - v) = x_K}$. Then (i.e. if the $p$ and $v$ could thus really be rendered **as zero**)**:** $X_P = X_J$ and $X_V = X_K$, and hence Eq. (A.1) will restore to the **ideal gas system** (IGS)**:** $(X_P \times X_V) = (X_J \times X_K) = RT = Y_T$.

Second, say that the $p$ and $v$ are, like the $X_P$ and $X_V$, assessed by ***physical*** measurements. Then the process of verifying the Boyle's law should, at the very first step, demand the replacement of the IGS by **a *four-variable* system** as Eq. (A.1). In fact, how the Boyle's law may for real gases be represented is unresolved. Nevertheless, several attempts were made to correct for the biases**:** $p$ and $v$. For example, if "$\boldsymbol{p = a/(X_V)^2}$" and "$\boldsymbol{v = b}$" (with "***a***" and "***b***" as the constants for a given gas [18]), then it is referred as the van der Waals system (VWS):

$$\left(X_P + {}^{a}/_{(X_V)^2}\right)(X_V - b) = RT = Y_T \qquad \text{(A.2)}$$

Eq. (A.2) explains why the systematic deviations as "*p*" and "*v*" are gas-specific, i.e. why, for any *given* temperature and pressure, *different* gases occupy *different volumes*. Further, unlike the IGS, the VWS can be shown to belong to the F.2 family, i.e.**:** $\boldsymbol{M_P^T = [X_P / (X_P + [a / (X_V)^2])]}$; and $\boldsymbol{M_V^T = [(X_P\, X_V + a\, [(2b / X_V) - 1] / X_V) / Y_T]}$. In other words, the uncertainty **(**$\varepsilon_T^{VWS}$, in an estimate of $Y_T$ *obtained by the Eq. (A.2)***)** would be governed by the given VWS. That is (though**:** $\boldsymbol{\varepsilon_T^{IGS}} = \varepsilon_T = (u_J + u_K) = \boldsymbol{f_T(u_J,u_K)}$, cf. above); $\boldsymbol{\varepsilon_T^{VWS}} = \sum_{i=1}^{2} |M_i^T|\, u_i = \boldsymbol{f_T(X_P,X_V,u_P,u_V)}$.

Now, say that the nitrogen gas ($a$ = 1.39 lit.$^2$.-atm, and $b$ = 0.0392 lit. [18]) at 273.16°K is an example of **perfect** VWS. That is, if**:** $X_P = X_P^{N_2}$ = **400** atm., then Eq. (**A.2**) predicts**:** $X_V = X_V^{N_2}$ =

**0.0731855** liter. These in turn imply: $\boldsymbol{M_P^T}$ = 0.61 and $\boldsymbol{M_V^T}$ = 1.37; and/ or: $\varepsilon_T^{VWS} = \varepsilon_T^{N_2}$ = (0.61$u_P$ + 1.37$u_V$), where $u_P$ and $u_V$ stand for the measurement-uncertainties. Further, say: $u_P$ = $u_V$ = 1%. Then: $y_T = y_T^{N_2}$ = ($Y_T \pm \varepsilon_T^{N_2}$) = (**22.414 ± 1.98%**). For example "$x_P^{N_2}$ = ($X_P^{N_2}$ – 1%) = ***396*** and $x_V^{N_2}$= ($X_V^{N_2}$ – 1%) = ***0.0724536***" can be seen to yield: $y_T = y_T^{N_2}$ = ***21.97*** = ($Y_T$ – ***1.98%***).

However the actual measurement [18], against: $x_P^{N_2} = X_P^{N_2}$ = **400** atm. ($\Delta_P$ = **0**), had yielded (cf. above): $x_V^{N_2}$ = **0.0703** liter ($\Delta_V$ = [($x_V^{N_2}/X_V^{N_2}$) –1] = **–3.94%**). Therefore (cf. the LHS of Eq. A.2): $y_T^{N_2}$ = **21.187** liter atm., and hence: $\delta_T^{N_2}$ = **–5.47%**. Thus, the error $\delta_T^{N_2}$ is ≈5 fold reduced from that (**25.5%**, cf. above) for using the IGS. Yet, "$\delta_T^{N_2}$" is far more high than to be expected for the "*a*" and "*b*" values to be absolutely accurate and/ or for the behavior of nitrogen to be exemplary of the VWS. Nonetheless, the error ($\delta_T^{N_2}$ = **–5.47%**) is accountable by the theory (Eq. 4): $\delta_T^{Theo}$ = ($M_P^T\ \Delta_P + M_V^T\ \Delta_V$) = (0.61 × **0)** + (1.37 × (**–3.94**)) = **–5.4%**. Thus, as shown here, we may mean the bias-corrections (***p = a*/(*X*$_V$*)***$^2$ and ***v = b***) to really be imperfect.

Actually, "*a*" and "*b*" are temperature-dependent [18]. And, for *a* and *b* to also be **variables**, Eq. 6 predicts (while: $M_P^T$ = **0.61**; and $M_V^T$ = **1.37**, see above): $\boldsymbol{M_a^T}$ = [$a$ (1 – [$b$ / $X_V$]) / ($X_V\ Y_T$)] = **0.39**; and $\boldsymbol{M_b^T}$ = –[$b$ ($X_P$ + [$a$ / $(X_V)^2$]) / $Y_T$] = **–1.15**. That is (like: $\boldsymbol{M_P^T}$), $\boldsymbol{M_a^T}$ <1. However, "$|\boldsymbol{M_b^T}|$" is >**1** (i.e. as: $\boldsymbol{M_V^T}$). Therefore, even an error in "*b*" should significantly affect the result ($y_T^{N_2}$). Thus, e.g. for: $u_P = u_a = u_V = u_b = u_i$, $\varepsilon_T^{VWS} = \varepsilon_T^{N_2}$ = ($|M_P^T|u_P + |M_a^T|u_a + |M_V^T|u_V + |M_b^T|u_b$) = (0.61 + 0.39 + 1.37 + 1.15)$u_i$= **3.52**$u_i$ (cf. Eq. 9). Or: "$\varepsilon_T^{N_2}/u_i$" = **3.52**, which is ≈2 times higher than that ("$\varepsilon_T^{N_2}/u_i$" = **1.98**, cf. above) for the **two**-variable VWS. That is the above observed error: $|\ \delta_T^{N_2}\ |$ = **5.47%** is better explicable by the present consideration. However, it is difficult to predict the errors in "*a*" and "*b*" [18], and hence to confirm the fact. Yet, in support, it could be added **e.g.** that "$x_P^{N_2}$ = ($X_P^{N_2}$ – **0.1%**) = 399.6; $a$ = ($a$ – **0.1%**) = 1.38861; $x_P^{N_2}$= ($X_P^{N_2}$ – **0.1%**) = 0.0731123; and $b$ = ($b$ + **0.1%**) = 0.0392392" yield**:** $y_T$ = $y_T^{N_2}$ = 22.335 = ($Y_T$ – **0.352%**).

We now consider a case of *pressure* ($X_J$) measurement. The pressure ($x_J^{CO_2}$), for 1 mole of $CO_2$ gas occupying a volume ($\boldsymbol{X_K^{CO_2}}$) of **0.381** liter at 313.16°K, was estimated [18] to be **50** atm. Therefore**:** $y_T^{CO_2}$ = ($x_J^{CO_2}$ × $x_K^{CO_2}$) = ($x_J^{CO_2}$ × $\boldsymbol{X_K^{CO_2}}$) = **19.05** liter-atm. i.e. the deviation from the IGS ($Y_T$= $RT$ = **25.696** liter-atm.) is again in this case very high (–25.9%).

However, if $CO_2$ ($a$ = 3.60 lit.$^2$.-atm, and $b$ = 0.0428 lit. [18]]) ***obeys*** the VWS **(**i.e. if**:** $Y_T$ = **25.696** liter-atm., and $X_V^{CO_2}$ = $X_K^{CO_2}$ = **0.381** liter (**$\Delta_V$ = 0)**), then Eq. (A.2) predicts**:** $X_P^{CO_2}$ = **51.18** atm. Further, such a case imply (cf. Eq. 6)**:** $M_P^T$ = 0.67, $M_a^T$ = 0.33, $M_V^T$ = 0.47, and $M_b^T$ = –0.13. Thus, if $X_P^{CO_2}$ and $X_V^{CO_2}$ should **only** be the measured variables, then**:** "$\varepsilon_T^{CO_2}/u_i$" = **1.14**. But, the ***four***-variable VVS imply**:** "$\varepsilon_T^{CO_2}/u_i$" = ***1.60***.

However, for [18]**:** $x_P^{CO_2}$ = **50** atm. (i.e.**: $\Delta_P$** = [($x_P^{CO_2}$ /$X_P^{CO_2}$) –1] = **–2.31%**) and $x_V^{CO_2}$ = $X_V^{CO_2}$ = **0.381** liter (**$\Delta_P$ = 0**), Eq. (A.2) gives**:** $y_T^{CO_2}$ = **25.30** (i.e. $\delta_T^{CO_2}$ = **–1.55%**). Clearly, "$\Delta_P$" alone can account for "$\delta_T^{CO_2}$" (cf. Eq. 4**:** $\delta_T^{Theo} = M_P^T\, \Delta_P$ = (0.67 × [–2.31]) = –1.55%), but is too high to be believed as of random origin. Thus, again, the behavior of a real gas ($|\delta_T^{CO_2}|$= **1.55%**) appeals to be accounted for by the SSR-specific uncertainty consideration as**:** ($\varepsilon_T^{CO_2}/u_i$) = ***1.60***.

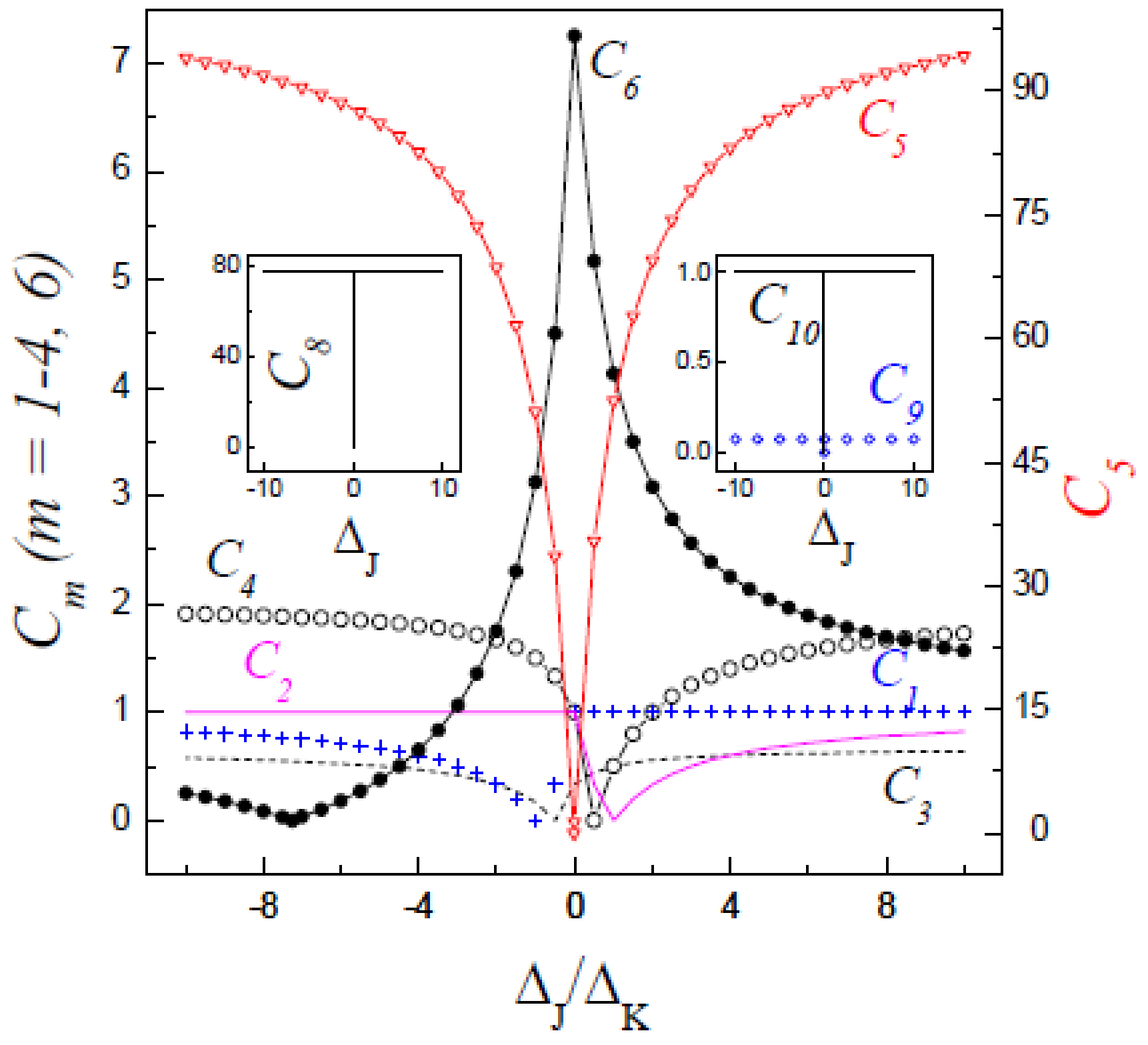

**Figure 1.** Predicted variations of $C_m$ ($m$ =*1-6*) vs. $\Delta_J/\Delta_K$, and $C_m$ ($m$ = *8-10*) vs. $\Delta_J$.

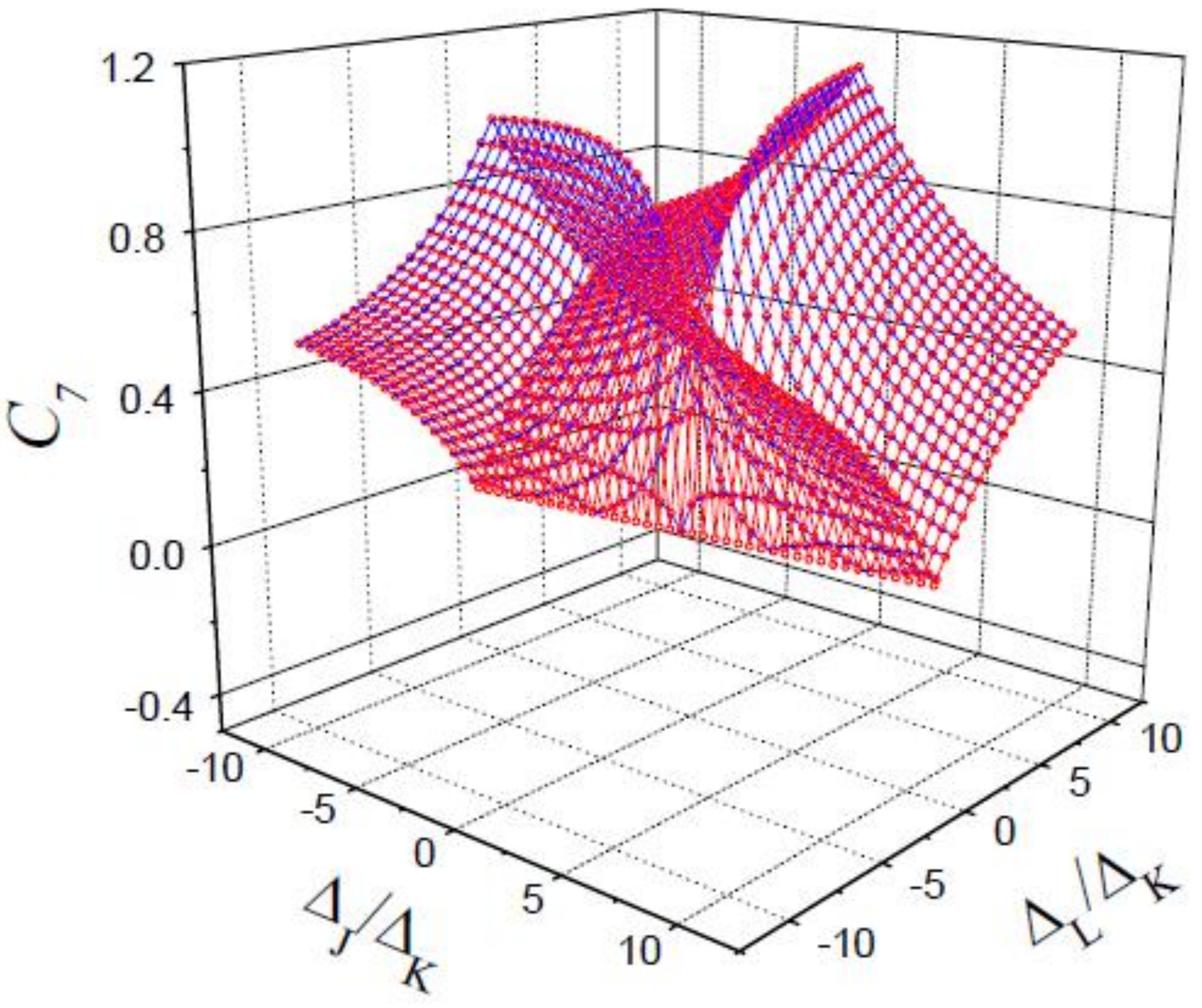

**Figure 2.** Variations of $C_7$ as a function of the error-ratios as $\Delta_J/\Delta_K$ and $\Delta_L/\Delta_K$. [The blue color refer to the variations in only the XZ planes (i.e. for fixed $\Delta_L/\Delta_K$ values), and the red color in the YZ planes (i.e. for fixed $\Delta_J/\Delta_K$ values)].